\documentclass[12pt]{article}
\usepackage{titlesec}
\titleformat{\section}{\bf\large}{\MakeUppercase \thesection.}{1em}{\MakeUppercase}
\usepackage{amssymb} 
\usepackage{amsmath}
\usepackage{amsthm}
\usepackage{mathtools}
\usepackage{dsfont}
\usepackage{bm}
\usepackage{subcaption}
\usepackage{enumerate}
\usepackage{graphicx}
\usepackage{color}
\usepackage{float}
\usepackage{natbib}
\usepackage{booktabs}
\usepackage{colortbl}
\usepackage{xcolor}
\usepackage{multirow}

\usepackage{xr}
\usepackage{xr-hyper}
\usepackage{refcount}

\usepackage{hyperref}
\usepackage{nameref}
\definecolor{TUMblue}{cmyk}{1, .54, .04, .19}
\hypersetup{
	colorlinks   = true, %Colours links instead of ugly boxes
	urlcolor     = TUMblue, %Colour for external hyperlinks
	linkcolor    = TUMblue, %Colour of internal links
	citecolor   = TUMblue %Colour of citations
}
\usepackage{cleveref}
\crefname{assumption}{}{}
\crefname{wexample}{weight estimator}{weight estimators}
\creflabelformat{assumption}{#2(A#1)#3}
\newtheorem{theorem}{Theorem}
\newtheorem{corollary}{Corollary}

\newtheorem{assumption}{Assumption}

\theoremstyle{remark}
\newtheorem{remark}{Remark}
\newtheorem{example}{Example}
\newtheorem{wexample}{Weight Estimator}

\usepackage{enumerate}
\usepackage{enumitem}
\newenvironment{subassumption}{\begin{enumerate}[label={(\roman*)},
			ref={\theassumption\roman*}]
		}{\end{enumerate}}
\allowdisplaybreaks
\newcommand{\E}{{\mathrm{E}}}

\newcommand{\R}{{\mathbb R}}

\newcommand{\Gcal}{{\mathcal G}}

\newcommand{\Xcal}{{\mathcal X}}
\newcommand{\Ycal}{{\mathcal Y}}

\newcommand{\bX}{{\bm X}}
\newcommand{\bY}{{\bm Y}}
\newcommand{\bZ}{{\bm Z}}
\newcommand{\bx}{{\bm x}}
\newcommand{\by}{{\bm y}}
\newcommand{\bz}{{\bm z}}
\newcommand{\bu}{{\bm u}}
\newcommand{\bv}{{\bm v}}

\newcommand{\bF}{{\bm F}}
\newcommand{\btheta}{{\bm \theta}}
\newcommand{\bmu}{{\bm \mu}}
\newcommand{\bSigma}{{\bm \Sigma}}

\newcommand{\bgam}{{\bm \gamma}}

\newcommand{\G}{\mathbb{G}}

\newcommand{\Ind}{\mathds{1}}
\newcommand{\var}{\mathrm{var}}
\newcommand{\cov}{\mathrm{cov}}
\newcommand{\Prob}{\mathrm{Pr}}

\newcommand{\wh}{\widehat}
\newcommand{\wt}{\widetilde}
\newcommand{\wb}[1]{\wh{\bm #1}}

\DeclareFontFamily{U}{mathx}{\hyphenchar\font45}
\DeclareFontShape{U}{mathx}{m}{n}{
	<5> <6> <7> <8> <9> <10>
	<10.95> <12> <14.4> <17.28> <20.74> <24.88>
	mathx10
}{}
\DeclareSymbolFont{mathx}{U}{mathx}{m}{n}
\DeclareFontSubstitution{U}{mathx}{m}{n}
\DeclareMathAccent{\widecheck}{0}{mathx}{"71}

\newcommand{\ethat}{\widehat{\eta}}
\newcommand{\etboot}{\widetilde{\eta}}
\newcommand{\thhat}{\widehat{\theta}}
\newcommand{\thtrue}{{\theta^{*}}}

\newcommand{\thtruet}{{\theta_t^{*}}}
\newcommand{\thboot}{\widetilde{\theta}}
\newcommand{\thboott}{{\thboot_t}}
\newcommand{\what}{\widehat{w}}
\newcommand{\wtrue}{{w^{*}}}
\newcommand{\wboot}{\widetilde{w}}

\newcommand{\gth}{g_{\theta}}
\newcommand{\gtht}{g_{\theta, t}}
\newcommand{\gtrue}{g_{\thtrue}}
\newcommand{\gtruet}{g_{\thtrue, t}}
\newcommand{\ghat}{g_{\thhat}}

\newcommand{\htht}{h_{n, \theta, t}}

\newcommand{\Vtrue}{V_{ \thtrue}}

\newcommand{\bias}{\beta_{n}}

\newcommand{\wn}{{w_n}}

\newcommand\independent{\protect\mathpalette{\protect\independenT}{\perp}}
\def\independenT#1#2{\mathrel{\rlap{$#1#2$}\mkern2mu{#1#2}}}

\makeatletter
\newcommand{\subalign}[1]{%
	\vcenter{%
		\Let@ \restore@math@cr \default@tag
		\baselineskip\fontdimen10 \scriptfont\tw@
		\advance\baselineskip\fontdimen12 \scriptfont\tw@
		\lineskip\thr@@\fontdimen8 \scriptfont\thr@@
		\lineskiplimit\lineskip
		\ialign{\hfil$\m@th\scriptstyle##$&$\m@th\scriptstyle{}##$\crcr
			#1\crcr
		}%
	}
}
\makeatother

\usepackage[most]{tcolorbox}
\definecolor{block-gray}{gray}{0.8}
\newtcolorbox{revquote}{colback=block-gray,grow to right by=-2mm,grow to left by=-2mm,boxrule=0pt,before skip=1em,after skip=1em,breakable}
\newtcolorbox{myquote}{colback=white,grow to right by=-2mm,grow to left by=-2mm,boxrule=1pt,before skip=1em,after skip=1em,breakable}

\newcommand{\blind}{1}
\newcommand{\supplement}{0}

\def\spacingset#1{\renewcommand{\baselinestretch}%
{#1}\small\normalsize} \spacingset{1}

\usepackage{pdfpages}

\begin{document}

\if1\blind
	{
		\if0\supplement
			{
				\title{\bf Solving Estimating Equations With Copulas}
			} \fi
		\if1\supplement
			{
				\title{\bf Solving Estimating Equations With Copulas: Online Supplement}
			} \fi

		\author{
			Thomas Nagler\thanks{
				Department of Statistics, Ludwig Maximilian University of Munich, Munich, Germany; email: mail@tnagler.com; ORCID: 0000-0003-1855-0046}   
			~and 
			Thibault Vatter\thanks{
			Meta, New York, USA; email: tvatter@fb.com; ORCID: 0000-0001-9212-0218} \\
		}
		\maketitle
	} \fi

\if0\blind
	{
		\bigskip
		\bigskip
		\bigskip
		\begin{center}
			\if0\supplement
				{
					{\LARGE\bf Solving Estimating Equations With Copulas}
				} \fi
			\if1\supplement
				{
					{\LARGE\bf Solving Estimating Equations With Copulas: Online Supplement}
				} \fi
		\end{center}
		\medskip
	} \fi

\if0\supplement
	{
		\bigskip
		\begin{abstract}
			Thanks to their ability to capture complex dependence structures, copulas are frequently used to glue random variables into a joint model with arbitrary marginal distributions.
			More recently, they have been applied to solve statistical learning problems
			such as regression or classification.
			Framing such approaches as solutions of estimating equations, we generalize them in a unified framework.
			We can then obtain simultaneous, coherent inferences across multiple regression-like problems.
			We derive consistency, asymptotic normality, and validity of the bootstrap for corresponding estimators.
			The conditions allow for both continuous and discrete data as well as parametric, nonparametric, and semiparametric estimators of the copula and marginal distributions.
			The versatility of this methodology is illustrated by several theoretical examples, a simulation study, and an application to financial portfolio allocation.
		\end{abstract}

		\noindent%
		{\it Keywords:} regression, quantile, nonparametric, statistical learning, bootstrap
		\vfill
	} \fi

\newpage
\spacingset{1.5} % DON'T change the spacing!
%!TEX root = ../copula_ee.tex

%TODO: what to do with the paper that tried to scoop us?
\section{Introduction}\label{sec:intro}

Any multivariate distribution is composed of marginal distributions, and a \emph{copula} characterizing dependence.
Copula-based models combine complex dependencies with arbitrary marginal distributions and have become increasingly popular over the last two decades.
They have also been applied to solve statistical learning problems like  mean regression \citep{Pitt2006,Kolev2009,Noh2013,Cooke2015,cai2018}, quantile regression \citep{Bouye2009,Chen2009, Noh2015,Kraus2017a,Remillard2017}, and classification \citep{Elidan2012, Han2013, Nagler2016, Carrera2019}.
And when parametric models fail \citep[e.g.,][]{Dette2014}, semi/nonparametric approaches can be used  \citep{Debacker2017,Schallhorn2017}.

%!TEX root = ../../copula_ee.tex

A criticism is that copula-based methods lead to overcomplicated inferential procedures and/or sub-optimal rates, as they use the joint distribution to extract features of the conditional distribution.
In other words, they solve a problem that is harder than necessary.
In this paper, we show that this has a flip side:
copula-based methods yield simultaneous and coherent inferences across arbitrary combinations of finite/infinite-dimensional and potentially constrained features of the conditional distribution.

%!TEX root = ../../copula_ee.tex

Consider for instance a portfolio manager tasked with investing in $d$ assets
conditionally on $p$ covariates representing the state of the economy.
Denote by $\bY\in \R^d$, $\bgam \in \R^d$, and $\bX \in \R^p$ the returns on the assets, fractions of total wealth invested in each asset, and covariates.
Quantitative portfolio management relies
on properties of the
distribution of $\bgam^{\top} \bY$ conditional on $\bX = \bx$,
like the expected return $\mu_{\bgam}$, standard deviation $\sigma_{\bgam}$, or a quantile $q_{\bgam}$, as functions of $\bgam$.
Because  $\mu_{\bgam}$ and $\sigma_{\bgam}$ are linear in the components of the conditional expectation vector and covariance matrix of $\bY$ given $\bX = \bx$, they only pose finite-dimensional problems.
The conditional quantile $q_{\bgam}$ is nonlinear in the portfolio weight $\bgam$, however, and therefore poses an infinite-dimensional problem.
The approach in this paper allows to construct estimators of $\{(\mu_{\bgam}, \sigma_{\bgam}, q_{\bgam})\colon \bgam \in \R^d\}$ with asymptotics holding uniformly over portfolio weights. Conveniently, positive definiteness of the estimated covariance matrix and monotonicity of quantiles are automatically preserved.

Broadly, we propose a framework to solve a wide range of statistical learning problems using copulas.
It is general enough to cover most types of regression, including mean, quantile, expectile, exponential family, or even instrumental variables and censored regression, as well as classification.
Such problems can be characterized by estimating equations involving conditional expectations.
Our approach builds on a key insight:
conditional expectations can be replaced by weighted unconditional ones, and the weight is a ratio of measures associated with copulas.

In \Cref{sec:eecop}, we construct corresponding estimators in two steps: first estimate the copula, and then solve an approximate version of the estimating equation.
Examples of compatible regression problems and estimators are given in \Cref{sec:examples}.
Given an estimated copula, all those problems can be solved simultaneously with  coherent answers.
We justify this approach by rigorous asymptotic theory in \Cref{sec:asymptotics}.
We prove consistency, weak convergence, and validity of a bootstrap procedure for the proposed estimators under verifiable assumptions.
Our asymptotic results are substantially more general than known results.
In particular, we allow for virtually all types of regression problems, continuous and discrete variables, and for parametric, semiparametric, and nonparametric estimators in a single framework.
In \Cref{sec:simulations}, we illustrate our method in simulated examples.
We revisit the portfolio performance and risk management example described above using real data in \Cref{sec:applications}.
\Cref{sec:conclusion} relates our results to the literature and outlines further applications to more involved regression problems.

An implementation of the methods of this paper is provided by the \texttt{R} package \texttt{eecop} \citep{nagler2020a}.
Proofs and additional results are in the supplementary material.

%!TEX root = ../copula_ee.tex

\section{Copula-based estimating equations} %%justplaying
\label{sec:eecop}

\subsection{Estimating equations for regression problems} \label{sec:eereg}

For two random vectors $\bY$ and $\bX$, denote by $F_{\bY, \bX}(\by,\bx) = \Prob(\bY \leq \bm  y, \bX \leq \bx)$ their joint distribution, $F_{\bY \mid \bX}(\by \mid \bx)  = \Prob(\bY \leq \by \mid \bX = \bx)$ the distribution of $\bY$ conditional on $\bX = \bx$, $F_{Y_j}(y_j) = \Prob(Y_j \leq y_j)$ and $F_{X_j}(x_j) = \Prob(X_j \leq x_j)$ the marginal distributions.

% TV: \Ycal and \Xcal are needed later in the text
Let $\bY \in \Ycal  \subseteq \R^d$ be the \emph{response} and $\bX \in \Xcal  \subseteq  \R^p$ a vector of \emph{covariates}.
The response is often univariate in the context of regression or classification, but can also be vector-valued as in the asset allocation example from \Cref{sec:intro}, or be enriched to encompass censoring indicators and instrumental variables (see \Cref{sec:conclusion}).

Fix $\bx \in \R^p$ and let the parameter of interest $\theta = \theta(\bx)$ be related to the conditional distribution $F_{\bY \mid \bX = \bx}$.
Denoting the parameter space by $\Theta$ and $\thtrue \in \Theta$ the true parameter, suppose there is a family of functions $\Gcal = \left\{ \gth:  \theta \in \Theta \right\}$  with the property
%Consider a collection of identifying functions $\{\gth \colon \theta \in \Theta\}$. 
\begin{align} \label{eq:ee_conditional}
	\E\bigl\{ \gtrue (\bY) \mid \bX= \bx \bigr\}  = 0.
\end{align}
Most regression problems can be formulated that way (see \Cref{sec:examples_identifying}).
The set $\Gcal$ is called a family of \emph{identifying functions} and~\eqref{eq:ee_conditional} the (population version of an) \emph{estimating equation}.
The name estimating equation stems from the fact that an estimator of $\thtrue$ can be constructed from solving a sample version of~\eqref{eq:ee_conditional}.
Unconditional expectations have a canonical sample version in the sample average.
Conditional expectations are more challenging,
but can be replaced by unconditional ones since $\E\bigl\{ \gtrue (\bY) \mid \bX= \bx \bigr\} = \E\bigl\{ \gtrue (\bY)  \wtrue (\bY) \bigr\}$ with
\begin{align} \label{eq:w_x}
	\wtrue(\by) = \frac{dF_{\bY \mid \bX}(\by \mid \bx)}{dF_\bY(\by)} =  \frac{dF_{\bY, \bX}(\by, \bx)}{dF_\bY(\by)dF_\bX(\bx)},
\end{align}
which can be understood as a weight function that accounts for the conditioning on $\bX = \bx$.
Hence, the estimating equation~\eqref{eq:ee_conditional} can be written equivalently as
\begin{align} \label{eq:ee_unconditional}
	\E\bigl\{ \gtrue (\bY)  \wtrue (\bY) \bigr\}  = 0,
\end{align}
which can be used to construct estimators.
More generally, \eqref{eq:ee_unconditional} holds for weights of the form
\begin{align*}
	\wtrue(\by) = \frac{dF_{\bY, \bX}(\by, \bx)}{dF_\bY(\by)dF_\bX(\bx)} \nu(\bx),
\end{align*}
with $\nu$ an arbitrary function such that $0 < |\nu(\bx)| < \infty$.
This can sometimes lead to useful simplifications as in \eqref{eq:wx_continuous} below.

\subsection{Representation using copulas} \label{sec:ee2cop}

Sklar's theorem \citep{Sklar1959} states that the joint distribution $F_\bZ$ of any random vector $\bZ \in \R^k$ can be represented as
\begin{align} \label{eq:sklar}
	F_\bZ(\bz) = C_\bZ\{F_{Z_1}(z_1), \dots, F_{Z_k}(z_k)\},
\end{align}
where the function $C_\bZ$ is called a \emph{copula}.
The copula is a distribution function with uniform margins and unique on the ranges of $F_{Z_i}$, $i \in \{1, \dots, d\}$.
Since, in what follows, it is evaluated solely on this set, potential non-uniqueness is not an issue.
The weight in \eqref{eq:w_x} can then be expressed as
\begin{align} \label{eq:weight}
	\wtrue(\by) = \frac{dC_{\bY, \bX}\{F_{Y_1}(y_1), \dots, F_{Y_d}(y_d), F_{X_1}(x_1), \dots, F_{X_p}(x_p)\}}{dC_{\bY}\{F_{Y_1}(y_1), \dots, F_{Y_d}(y_d))dC_{\bX}\{F_{X_1}(x_1), \dots, F_{X_p}(x_p))\}}.
\end{align}
The measure $dC_\bZ\{F_{Z_1}(z_1), \dots, F_{Z_k}(z_k)\}$ can be represented by a density with respect to the product of Lebesgue and counting measures for continuous and discrete variables respectively.
Suppose that $Z_1, \dots, Z_m$ are integer-valued and $Z_{m + 1}, \dots, Z_k$ are continuous, which generalizes to categorical variables by identifying ordered categories with integers and unordered categories with binary dummy variables.
Then
\begin{align*}
	dC_\bZ\{F_{Z_1}(z_1), \dots, F_{Z_k}(z_k)\}
	= \sum_{\mathclap{\subalign{(j_1, \dots, j_m) & \in \{0, 1\}^m \\ (j_{m + 1}, \dots, j_k) &= 0}}} (-1)^{\sum_{r = 1}^k j_r} \frac{ \partial^{k - m}C_\bZ\{F_{Z_1}(z_1 - j_1), \dots, F_{Z_k}(z_k - j_k)\} }{\partial z_{m +1} \cdots   \partial z_{k}} .
\end{align*}

In many relevant cases, the expression \eqref{eq:weight} for $\wtrue$ can be simplified further.
% \begin{remark}[Continuous random vectors]
For instance, copula models are most commonly applied to continuous random vectors.
If $\bZ$ is a continuous random vector with joint density $f_\bZ$ and marginal densities $f_{Z_k}$, $k = 1, \dots, d$, we can take derivatives in~\eqref{eq:sklar} to obtain
\begin{align*}
	f_\bZ(\bz) = c_\bZ\{F_{Z_1}(z_1), \dots, F_{Z_d}(z_d)\} \times \prod_{k = 1}^d f_{Z_k}(z_k),
\end{align*}
with $c_\bZ$ the density corresponding to $C_\bZ$.
Hence,~\eqref{eq:ee_conditional} is equivalent to~\eqref{eq:ee_unconditional} with
\begin{align} \label{eq:wx_continuous}
	\wtrue(\by) = \frac{c_{\bY, \bX}\{F_{Y_1}(y_1), \dots, F_{Y_d}(y_d), F_{X_1}(x_1), \dots, F_{X_p}(x_p)\}}{c_{\bY}\{F_{Y_1}(y_1), \dots, F_{Y_d}(y_d))},
\end{align}
where we dropped $c_{\bX}$ as it does not depend on $\by$ or $\theta$.
And because copulas have uniform marginals, $d = 1$ implies $c_{Y} \equiv 1$ and $\wtrue(y) = c_{Y, \bX} \{F_{Y}(y), F_{X_1}(x_1), \dots, F_{X_p}(x_p)\}.$
% \end{remark}

% TV: Is this necessary given that our classification example uses Baye's rule and we never actually do anything with discrete responses?

% \begin{remark}[Discrete response]
% 	If $Y \in \Z$ and $\bX \in \R^p$ is continuous, we have  $f_Y(y) = \Pr(Y = y)$.
% 	Defining $\dot C_{Y, \bX}(u, v_1, \dots, v_p) = \partial^p C(u, v_1, \dots, v_p) / \prod_{k = 1}^p \partial v_k$, \eqref{eq:ee_conditional} is equivalent to~\eqref{eq:ee_unconditional} with
% 	\begin{align*}
% 		\wtrue(y) =
% 		\frac{\dot  C_{Y, \bX}\{F_{Y}(y), F_{X_1}(x_1), \dots, F_{X_p}(x_p)\} - \dot C_{Y, \bX}\{F_{Y}(y - 1), F_{X_1}(x_1), \dots, F_{X_p}(x_p)\}}
% 		{f_Y(y)}.
% 	\end{align*}
% \end{remark}

\subsection{Estimators for copula-based estimating equations}\label{sec:eecop2}

Suppose we observe an \emph{iid} sequence of random vectors $(\bY_1, \bX_1), \dots, (\bY_n, \bX_n)$.
We can use a sample version of~\eqref{eq:ee_unconditional} to construct estimators for the parameter $\thtrue$.
To do this, all unknown quantities in the unconditional estimating equation are replaced by estimates.

% TV: later, we write the two steps as first estimate the copula and 
% then replace the population expectation by sample average.

Let $\what(\by) = \what(\by; \bY_1, \bX_1, \dots,$ $\bY_n, \bX_n)$ be an estimator of $\wtrue$ (see \Cref{sec:examples_copula} for examples).
Recall that $\wtrue$ is a ratio of measures associated with copulas.
Hence, we can construct $\what$ by plugging in estimators of the copulas and margins.
This allows us to harness the rich toolbox of existing copula models and associated estimating techniques.

Then we define an estimator $\thhat = \thhat(\bY_1, \bX_1, \dots, \bY_n, \bX_n)$ of $\thtrue$ as the solution to
\begin{align} \label{eq:ee_estimator}
	\frac 1 n \sum_{i = 1}^n g_{\thhat} (\bY_i) \what (\bY_i) = 0,
\end{align}
using the sample average as a natural estimate of the expectation in~\eqref{eq:ee_unconditional}.
For a few specific copula models, the expectation $\E\{g_{\thhat} (\bY) \what (\bY)\}$ might be available in closed form.
Or it could be computed using numerical integration if, additionally, an estimate of the density $f_\bY$ is available.
But the sample average provides a simpler and generally valid method.

\section{Examples} \label{sec:examples}

The procedure described in \Cref{sec:eecop2} is quite versatile.
With different choices of the identifying function $\gth$, we can estimate various features of the conditional distribution.
In the following, we introduce popular examples covered by the theory developed in \Cref{sec:asymptotics}.

\subsection{Examples of identifying functions} \label{sec:examples_identifying}
\begin{example}[Mean regression] \label{ex:mreg}
	A classical example is $\thtrue = \E(Y \mid \bX= \bx)$ and $\gth(y) = y - \theta$.
	Given an estimator $\what$ of the weight $\wtrue$, the estimating equation~\eqref{eq:ee_estimator} has the explicit solution
		$\widehat{\theta} = \sum^n_{i=1} Y_i \, \what(Y_i)/\sum^n_{i=1} \what(Y_i)$.
	It is similar to the Nadaraya-Watson estimator for the conditional mean,
	albeit the weights are also functions of the response $Y$.
\end{example}

\begin{example}[Quantile regression] \label{ex:qreg}
	Let $\thtruet = F^{-1}_{Y \mid \bX}(t \mid \bX = \bx)$ be the conditional $t$-quantile at level $t \in T = (0,1)$
	and consider all levels jointly.
	The parameter of interest is the conditional quantile function $\thtrue = \{\thtruet\colon t \in T\}$
	and $\Theta$ is a space of functions from $T$ to $\R$.
	Then solving~\eqref{eq:ee_unconditional} with  $\gtht(y) = t  - \Ind(y < \theta_t)$
	identifies $\thtrue$.
	Here, the identifying function $\gth = \{\gtht\colon t \in T\}$ is also indexed by the quantile level $t$.
	% The corresponding sample version~\eqref{eq:ee_estimator} can only be solved numerically, although there are efficient algorithms to compute ``weighted quantiles''.
	%Note that \Cref{eq:ee_estimator} is asymptotically equivalent to the estimator of \cite{Noh2013}, where the denominator is $n c_{X}\{F_{X_1}(x_1), \dots, F_{X_p}(x_p)\}$. However, our denominator, a sum over quantities already calculated for the numerator, is computationally more efficient. \qed
\end{example}

\begin{example}[Expectile regression] \label{ex:expectile}
	Expectiles generalize the mean of a distribution similarly as to how quantiles generalize the median \citep[e.g.,][]{Newey1987}.
	They are also indexed by a level $t \in (0, 1)$, where $t = 1/2$ corresponds to mean regression.
	The conditional expectile is identified by~\eqref{eq:ee_unconditional} with
		$\gtht(y) = t(y - \theta_t) \Ind(y \ge \theta_t) - (1 - t)  (\theta_t -y)\Ind(y <   \theta_t)$.
\end{example}

\begin{example}[Exponential family regression] \label{ex:exp_fam}
	Suppose $f_{Y\mid \bX = \bx}$ is a one-parameter exponential family with canonical parameter $\theta$, that is $f(y;\theta) = h(y)\exp \left\{  a(y)\theta - b(\theta) \right\}$
	where $h$, $a$, and $b$ are known functions.
	Using the score equations, $\theta$ can be identified via
	$\gth(y) = a(y) - b^\prime(\theta)$.
\end{example}

\begin{example}[Binary classification] \label{ex:class}
	Let $Y \sim \mathrm{Bernoulli}(p)$ be a class indicator with
	the target being the conditional probability $\thtrue = P( Y = 1 \mid \bX = \bx) = \E(Y \mid \bX = \bx)$, a special case of mean regression.
	Since $F_{\bX \mid Y}(\bx \mid y) = F_{\bX \mid Y = 0}(\bx)\Ind(y = 0) + F_{\bX \mid Y = 1}(\bx)\Ind(y = 1)$, Bayes' rule leads to
	\begin{align*}
		\wtrue(y)
		%\frac{dF_{Y, \bX}(y, \bx)}{dF_Y(y)dF_\bX(\bx)} 
		= \frac{dF_{\bX \mid Y}(\bx\mid y)}{dF_\bX(\bx)}
		%= \frac{dF_{\bX \mid Y}(\bx\mid y)}{p dF_{\bX \mid Y = 0}(\bx) +(1 - p) dF_{\bX \mid Y = 0}(\bx)}.  
		= \frac{dF_{\bX \mid Y = 0}(\bx)\Ind(y = 0) + dF_{\bX \mid Y = 1}(\bx)\Ind(y = 1)}{(1-p) dF_{\bX \mid Y = 0}(\bx) + p dF_{\bX \mid Y = 0}(\bx)} .
	\end{align*}
	Replacing all quantities by estimates and solving~\eqref{eq:ee_estimator} with $\gth(y) = y - \theta$ yields
	\begin{align*}
		\thhat =  \frac{\widehat p \wh{d F}_{\bX \mid Y = 1}(\bx)}{(1-\wh p) \wh{dF}_{\bX \mid Y = 0}(\bx) + \wh p \wh{dF}_{\bX \mid Y = 1}(\bx)},
	\end{align*}
	where $\wh p = n^{-1} \sum_{i = 1} \Ind(Y_i = 1)$.
	Modeling $d F_{\bX \mid Y=0}$ and $d F_{\bX \mid Y=1}$ with copulas and marginal distributions,
	such classifiers have been used in \cite{Elidan2012, Nagler2016, Carrera2019}, but so far without asymptotic guarantees.
\end{example}

\subsection{Estimators for the weight function}
\label{sec:examples_copula}

In this section, we discuss a few simple estimators of $\wtrue$.
We focus on continuous data to simplify our exposition.

\begin{wexample}[Fully parametric estimator] \label{wex:p}
	Let $\bm \eta_\bY$, $\bm \eta_\bX$ and $\bm \eta_C$ be parameter vectors, indexing families of marginal and copula densities, and $\bm \eta = (\bm \eta_\bY, \bm \eta_\bX, \bm \eta_C)$.
	This defines a parametric model for the weight \eqref{eq:wx_continuous}:
	\begin{align*}
		w(\by; \bm \eta) = \frac{c_{\bY, \bX}\{F_{Y_1}(y_1; {\bm \eta}_{Y_1}), \dots, F_{Y_d}(y_d; {\bm \eta}_{Y_d}), F_{X_1}(x_1; {\bm \eta}_{X_1}), \dots,  F_{X_d}(x_p; {\bm \eta}_{X_p}); \bm \eta_C\}}{c_{\bY}\{F_{Y_1}(y_1; {\bm \eta}_{Y_1}), \dots, F_{Y_d}(y_d; {\bm \eta}_{Y_d});  \bm \eta_C\}}.
	\end{align*}
	If $\bm \ethat$ is an estimator for the true parameter $\bm \eta^*$, such as a maximum-likelihood or method of moment estimator, the estimated weight is then simply $\what (\by) = w(\by; \bm \ethat)$.
	% We further write $\bm \ethat$ for a maximum-likelihood estimator of the true $\bm \eta^*$:
	% \begin{align*}
	%   \bm \ethat &= \arg\max_{\bm \eta} \sum_{i = 1}^n \log f_{Y \bX}(Y_i, \bX_i; \bm \eta),
	% \end{align*}
	% where $f_{\bm Y, \bm X}(y, \bm x)$ is the joint density of $(Y, \bm X)$.
	% The estimated weight is then simply
	% \begin{align*}
	%   \what (y) =dC_{Y \bX}\{F_Y(y; \widehat{\bm \eta}_Y), F_{X_1}(x_1; \widehat{\bm \eta}_{X_1}), \dots,  F_{X_d}(x_d; \widehat{\bm \eta}_{X_d}); \widehat{\bm \eta}\}.
	% \end{align*}
\end{wexample}

\begin{wexample}[Semiparametric estimator] \label{wex:s}
	Semiparametric copula models combine a parametric model for the copula density $c_{\bY, \bX}(\cdot; \bm \eta)$ with nonparametric margins.
	Denote by $\bm \ethat$ an estimator of the true copula parameter $\bm \eta^*$ in such a semiparametric model.
	Examples for such estimators are the pseudo-maximum-likelihood estimator of \citet{Genest1995} or the method-of-moment type estimators discussed in \citet{tsukahara2005}.
	A semiparametric estimator for the weight \eqref{eq:wx_continuous} is then given by
	\begin{align*}
		\what(\by) = \frac{c_{\bY, \bX}\{\wh  F_{Y_1}(y_1), \dots, \wh  F_{Y_d}(y_d), \wh  F_{X_1}(x_1), \dots,  \wh  F_{X_p}(x_p); \wh{\bm \eta}\}}{c_{\bY}\{\wh  F_{Y_1}(y_1), \dots, \wh F_{Y_d}(y_d);  \wh{\bm \eta}\}},
	\end{align*}
	where $\wh F$ denotes the empirical distribution function.
\end{wexample}

\begin{wexample}[Simple kernel estimator]\label{wex:n}
	Observe that taking
	\begin{align*}
		\wtrue(\by) = \frac{c_{\bY, \bX}\{F_{Y_1}(y_1), \dots, F_{Y_d}(y_d), F_{X_1}(x_1), \dots, F_{X_p}(x_p)\}}{c_{\bY}\{F_{Y_1}(y_1), \dots, F_{Y_d}(y_d)\}} \prod_{j = 1}^p  f_{X_k}(x_k) = \frac{f_{\bY, \bX}(\by, \bx)}{f_\bY(\by)}
	\end{align*}
	as weight function solves~\eqref{eq:ee_unconditional}.
	Then a natural estimator is
	$\what(\by) = \wh f_{\bY, \bX}(\by, \bx)/\wh f_\bY(\by)$,
	where $\wh f_{\bY, \bX}$ and $\wh f_{\bY}$ are estimators of $f_{\bY, \bX}$ and $f_{\bY}$.
	As an example, consider kernel density estimators (KDEs).
	For some univariate probability density $K\colon \R \to \R_{\ge 0}$ and bandwidth sequences $b_n, \sigma_n \to 0$, define the KDE for $f_{\bY, \bX}$ as
	\begin{align*}
		\wh f_{\bY, \bm X}(\by, \bx) = \frac{1}{nb_n^d \sigma_n^p}\sum_{i=1}^n \prod_{j = 1}^d K\biggr(\frac{Y_{i, j} - y_j}{b_n}\biggr)\prod_{k = 1}^p K\biggl(\frac{X_{i, k} - x_k}{\sigma_n}\biggr).
	\end{align*}
	Its margin  $
		\wh f_{\bY}(\by) = \int \wh f_{\bY, \bm X}(\by, \bx) d\bx =  n^{-1} b_n^{-d} \sum_{i = 1}^n \prod_{j = 1}^d  K\{(Y_{i, j} - y_j)/b_n\}$
	is also a KDE.
\end{wexample}
 
%!TEX root = ../copula_ee.tex

\section{Asymptotic theory}
\label{sec:asymptotics}

In this section, we derive some asymptotic properties of $\thhat$.
\Cref{sec:setup_notation} introduces required notations.
For our main results in \Cref{sec:main_results}, we use a general framework to encompass a wide range of identifying functions and weight estimators.
Then, \Cref{sec:corollaries} contains specialized results corresponding to the parametric, semiparametric, and nonparametric weight estimators of \Cref{sec:examples_copula}.
The assumptions are stated and discussed in \Cref{sec:asm}.
The proofs and additional verifications of the assumptions for the examples from \Cref{sec:examples_identifying} are in the supplementary material.

\subsection{Setup and notation}\label{sec:setup_notation}

We use $\to_P$ and $o_P\bigl(a_n\bigr)$ for convergence in probability without rate and with rate $a_n$, and $\leadsto$ for weak convergence.
For an arbitrary set $T$, denote the space of all bounded functions from $T$ to $\R$ by $\ell^\infty(T) = \{f\colon T \to \R, \left\| f \right\|_{T} < \infty\}$, with $\left\| f \right\|_{T} = \sup_{t \in T}\vert f(t) \vert$.

%!TEX root = ../copula_ee.tex

Assume that the parameter space satisfies $\Theta \subseteq \ell^\infty(T)$ for some indexing set $T$.
For simplicity, assume $T$ to be a compact subset of a Euclidean space.
It means that any $\theta \in \Theta$ and $\gth \in \Gcal = \{g_{\theta} \colon \theta \in \Theta \}$ is indexed by $T$, that is $\theta = \{\theta_t \colon t \in T\}$ and $g_{\theta} = \{g_{\theta, t} \colon t \in T\}$.
In particular, this holds for the estimator $\thhat$, true parameter $\thtrue$, and the corresponding $\ghat$ and $\gtrue$.
For scalar parameters of interest, we may take $T = \{ 1 \}$, implying that $\Theta$ is isometric to $\R$, and write $\theta \in \Theta \subseteq \R$ by slight abuse of notation.
As for vectors of dimension $k$, one can use $T=\{1, \dots, k\}$ with $\theta_t$ referring to the corresponding vector's $t$-th component, translating into $\btheta \in \Theta \subseteq \R^k$ and $\|\btheta \|_{T} = \max_{t=1, \dots, K} |\theta_t | $ by the same abuse of notation.
An example of the general case, where the parameter of interest is a genuine function from $T$ to $\R$, is quantile regression (see \Cref{ex:qreg}). Here $T=(0,1)$ is a natural choice with $\Theta$ being a space of functions from $(0,1)$ to $\R$ and $\theta \in \Theta$ being indexed by the quantile level.

Recall that $\thhat = \thhat(\bx)$ and $\thtrue = \thtrue(\bx)$ are also functions of the value conditioned upon through $\bX = \bx$.
This is reflected in their definitions~\eqref{eq:ee_estimator} and~\eqref{eq:ee_unconditional} through $\what$ and $\wtrue$.
But $\thhat$ as a function of $\bx$ cannot have a meaningful limit in many cases of practical interest.
Hence, we assume $\bx$ fixed for the remainder of this section and simply write $\thhat$ and $\thtrue$.

\subsection{Main results}\label{sec:main_results}

Our first result shows that $\thhat$ is consistent, uniformly in the indexing set $T$.
% TV: already stated above
% For a detailed list and brief discussion of the assumptions, see \Cref{sec:asm}.
\begin{theorem}[Consistency]
	\label{thm:consistency}
	Under \Crefrange{asm_new:identifiability}{asm_new:h_functions},
	$ \| \thhat -\thtrue \|_T \to_P 0$ as $n \to \infty$.
\end{theorem}

If the parameter of interest is scalar, vector, or matrix valued, uniform consistency is the usual consistency, due to the problem's finite dimensional nature.
But \Cref{thm:consistency} is insufficient for statistical inference.
For instance, to test hypotheses and construct confidence bands, an asymptotic distribution is needed.
And to establish weak convergence, we need to specify the estimator $\what$ further.

In what follows, we assume that $\what$ is asymptotically linear in the sense that there is a sequence of  functions $\wn\colon \bm \R^{2d + p} \to \R$ and a sequence $r_n^{-1} = o(n^{1/2})$ such that
\begin{align} \label{eq:wapprox1}
	\what(\by) \approx \frac{1}{n} \sum_{i=1}^n \wn(\by,\bY_i,\bX_i) + o_P\bigl(n^{-1/2}r_n\bigr),
\end{align}
in a sense that is made precise by \Cref{asm_new:wn}.
This assumption is satisfied by many estimators of $\wtrue$ including those from \Cref{sec:examples_copula}, see \Cref{sec:corollaries}.
The rate $r_n$ allows to encompass both parametric and nonparametric estimators of $\wtrue$.
While $r_n$ is a diverging sequence for nonparametric estimators, $r_n = 1$ gives the standard $\sqrt{n}$ rate for parametric ones.

\begin{theorem}[Weak convergence]
	\label{thm:weak_convergence}
	Under \Crefrange{asm_new:identifiability}{asm_new:frechet},
	\begin{align*}
		r_n^{-1}\sqrt{n}(\thhat - \thtrue - \bias) \leadsto -\Vtrue^{-1}\G \mbox{ in } \ell^\infty(T),
	\end{align*}
	where $\Vtrue$ is the Fr\'echet derivative of the map $\theta \mapsto \E \bigl\{ \gth(\bY) \wtrue(\bY) \bigr\}$ taken at $\thtrue$,
	\begin{align*}
		\bias = -\Vtrue^{-1}\E\{\gtrue(\bY)w_n(\bY, \bY', \bX')\}
	\end{align*}
	for $(\bY', \bX')$ an independent copy of $(\bY, \bX)$,
	and $\G$ is a tight, mean-zero Gaussian with
	\begin{align*}
		&\mathrm{cov}(\G_{t_1}, \G_{t_2}) = \lim_{n \to \infty}r_n^{-2} \E\{h_{n,t_1}(\bY',\bX')  h_{n,t_2}(\bY',\bX')\},
	\end{align*}
where $h_{n, t}(\by',\bx') =\E\{\gtruet(\bY)w_n(\bY, \by', \bx') + \gtruet(\by')w_n(\by', \bY, \bX)\}$ for  $(\by', \bx') \in \Ycal \times \Xcal$.
\end{theorem}

While this result is functional, it can be understood in the usual sense for scalar and vector valued parameters of interest.
For scalars, we can take $T = \{1 \}$, so $\G$ is mean-zero univariate Gaussian.
For $k$-dimensional vectors, using $T=\left\{1, \dots, k \right\}$, $\Vtrue$ is the derivative of a map from $\R^k$ to $\R^k$, that is a $k \times k$ matrix.
And $\G$ is a mean-zero $k$-dimensional Gaussian with covariance matrix with entries $\cov(\G_i, \G_j)$ for $1 \leq i,j \leq k$.

To better understand \Cref{thm:weak_convergence}, recall that $\thhat$ is a plug-in type estimator.
Its asymptotic distribution combines the effects of two steps: replacing
(i) $\wtrue$ with $\what$, and (ii) a population expectation by a sample average.
Since (ii) is unbiased, the bias $\bias$ is driven by the bias of $\what$.
This becomes obvious when writing
\begin{align*}
	\bias = -\Vtrue^{-1}\E\{\gtrue(\bY)w_n(\bY, \bY', \bX')\}= - \Vtrue^{-1}\E\{g_{ \thtrue}(\bY)b_{n}(\bY)\},
\end{align*}
with $b_{n}(\by) = \E[\wn(\by, \bY', \bX')] - \wtrue (\by)$ the first order term in the bias of $\what$.
Hence, $\bias$ is proportional to a weighted average of the bias of $\what$, where the averaging accounts for the second step.
For interpreting the variance, assume that $b_{n}(\by) = o(1)$ such that
\begin{align*}
	h_{n, t}(\by', \bx') \approx \E\{\gtruet(\bY)w_n(\bY, \by', \bx')\} + \gtruet(\by')\wtrue(\by').
\end{align*}
Then
\begin{align*}
	\var\{h_{n,t}(\bY',\bX')\} & \approx  	\mathrm{var} \bigl[\E\{\gtrue(\bY)w_n(\bY, \bY', \bX') \mid \bY', \bX' \} ] + \mathrm{var}\{ \gtruet(\bY)\wtrue(\bY)\} \\
	                           & + 2 \bigl[ \E\{\gtrue(\bY)w_n(\bY, \bY', \bX') \mid \bY', \bX' \} \gtruet(\bY')\wtrue(\bY') ].
\end{align*}
The first and second terms respectively reflect the variability caused by steps (i) and (ii).
And the third echoes the dependence between both steps, since the same data is used twice.

\Cref{thm:weak_convergence} allows to compute the limiting distribution for specific choices of identifying functions and weight estimators.
But such computations can be rather involved in practice, especially for complex $\what$.
To remedy this issue, we propose a \emph{bootstrap} method.
The idea is to define a new estimator $\thboot$ based on a randomly reweighted version of the data.
The distribution of $\sqrt{n}(\thhat - \thtrue)$ is then approximated by that of $\sqrt{n}(\thboot - \thhat)$.

Specifically, let $\xi_1, \dots, \xi_n$ be an \emph{iid} sequence of positive random variables independent of the data and satisfying $\mathrm{E}(\xi_1) = \mathrm{var}(\xi_1) =  1$, $\mathrm{E}(\vert \xi_1\vert^{2+\epsilon}) < \infty$ for some $\epsilon > 0$.
For instance, $\xi_i\sim \mathrm{Exp}(1)$ is the \emph{Bayesian bootstrap} of \cite{Rubin1981}.
Define the bootstrap estimator $\thboot = \bigl\{ \thboott: t \in T \bigr\}$ as solving
\begin{align*}
	\frac 1 n \sum_{i = 1}^n \xi_i g_{\thboot, t}(\bY_i)\wboot(\bY_i) = 0,
\end{align*}
where the bootstrapped weight $\wboot$ is constructed so that
\begin{align}  \label{eq:wapprox2}
	\wboot (\by) \approx  \frac 1 n \sum_{i=1}^n \xi_i \wn(\by, \bY_i, \bX_i) + o_p(n^{-1/2}r_n),
\end{align}
in a sense that is made precise by \Cref{asm_new:boot}. 
In \Cref{sec:corollaries}, we explain how such bootstrapped weights are constructed for the \cref{wex:p,wex:s,wex:n}.

Our final theorem implies that  $r_n^{-1}\sqrt{n}(\thboot - \thhat)$ converges to the same limit as in \Cref{thm:weak_convergence}. 
See also \citet{bucher2019note} for equivalent formulations of bootstrap validity.
\begin{theorem}[Validity of the bootstrap]  \label{thm:bootstrap}
	Under \Crefrange{asm_new:identifiability}{asm_new:boot},
	\begin{align*}
		r_n^{-1}\sqrt{n}(\thhat - \thtrue - \beta_n, \thboot - \thhat) \leadsto -(\Vtrue^{-1} \G, \Vtrue^{-1} \wt \G) \quad \mbox{ in } \ell^\infty(T) \times \ell^\infty(T),
	\end{align*}
	where $\wt \G$ is an independent copy of $\mathbb G$ and
	$\Vtrue, \G$  are as in \Cref{thm:weak_convergence}.
\end{theorem}

\begin{remark}
	The resampling technique of \cite{Efron1979} uses dependent bootstrap weights $(\xi_1, \dots, \xi_n) \sim \mathrm{Multinomial}(n, 1/n, \dots, 1/n)$.
	% This can be resolved by a Poissonization argument, albeit with substantial additional effort.
	Our formulation simplifies the asymptotic analysis and is rather natural in the context of estimating equations.
\end{remark}

\subsection{Examples continued} \label{sec:corollaries}
\Cref{thm:consistency,thm:weak_convergence,thm:bootstrap} are general enough to cover a broad range of estimation methods and regression problems.
In particular, they apply to all the examples given in \Cref{sec:examples}.
In this section, we give three corollaries corresponding to the \cref{wex:p,wex:s,wex:n}.
% Focussing on specific estimation methods lead to simpler conditions given in \Cref{sec:asm:cor}. 
% For \cref{wex:p,wex:s,wex:n}, \Cref{asm_new:p,asm_new:s,asm_new:n} respectively allow to replace \Cref{asm_new:wn,asm_new:g_wn,asm_new:consistency,asm_new:h_functions,asm_new:h_functions_cont,asm_new:boot}.

The parametric \cref{wex:p} is defined as $\what(\cdot) = w(\cdot; \wb \eta)$, where $\wb \eta$ is the estimated model parameter.
Assume that $\bm \ethat- \bm \eta^* = n^{-1} \sum_{i=1}^n \bm \gamma(\bY_i, \bX_i)  + o_P(n^{-1/2})$  for some function $\bm \gamma$	with $\E\{\bm \gamma(\bY, \bX)\} = 0$, which is satisfied for both maximum likelihood and method of moments estimators under some regularity conditions.
If $w$ is sufficiently smooth, we get
\begin{align*}
	\what(\by) \approx w(\by; \bm \eta^*) + \frac 1 n \sum_{i = 1}^n \nabla_{\bm \eta}^\top w(\by; \bm \eta^*)  \bm \gamma(\bY_i, \bX_i) + o_p(n^{-1/2}).
\end{align*}
In other words, \eqref{eq:wapprox1} holds with $r_n=1$ and $w_n(\by, \bY_i, \bX_i) =  w(\by; \bm \eta^*) +  \nabla_{\bm \eta}^\top w(\by; \bm \eta^*)  \bm \gamma(\bY_i, \bX_i)$.
Furthermore, \eqref{eq:wapprox2} holds with the same $r_n$ and $w_n$ if the bootstrap estimator $\bm \etboot$ satisfies $\bm \etboot - \bm \eta^* = n^{-1} \sum_{i=1}^n \xi_i \bm \gamma(\bY_i, \bX_i)  + o_P(n^{-1/2})$.
For example, if $\wb \eta$ is the maximum likelihood estimator, we may take  $\bm \etboot = \arg\max_{\bm \eta} \sum_{i = 1}^n \xi_i \log f_{\bY, \bX}(\bY_i, \bX_i; \bm \eta).$

\begin{corollary}[Fully parametric estimator] \label{cor:p}
	Under \Cref{asm_new:identifiability,asm_new:euclidean,asm_new:frechet,asm_new:p}, the \cref{wex:p} satisfies the conditions of \Cref{thm:consistency,thm:weak_convergence,thm:bootstrap}.
	The bias and covariance are given by $\beta_n = 0$ and  $\cov(\G_{t_1}, \G_{t_1})  = \cov(Z_{t_1}, Z_{t_2})$, where
	\begin{align*}
		Z_t = \gtruet(\bY) w(\bY; \bm \eta^*) +  \nabla_{\bm \eta^*}^\top \E\{\gtruet(\bY)  w(\bY; \bm \eta^*) \} \bm \gamma(\bY, \bX).
	\end{align*}
\end{corollary}
\noindent A similar result was obtained by \citet{Remillard2017} for the weak convergence of conditional quantile estimators.
For the semiparametric \cref{wex:s}, we can proceed similarly, but the resulting variance is larger due to nonparametric margins estimation.
% TV: \omega wasn't used anywhere in the main text and we define it in the asumption anyway
% TN: Should also suffice to have the ugly expression for lambda just in the  assumption

\begin{corollary}[Semiparametric estimator] \label{cor:s}
	Under \Cref{asm_new:identifiability,asm_new:euclidean,asm_new:frechet,asm_new:s}, the \cref{wex:s} satisfies the conditions of \Cref{thm:consistency,thm:weak_convergence,thm:bootstrap}.
	The bias and covariance are given by $\beta_n = 0$ and  $\cov(\G_{t_1}, \G_{t_1})  = \cov(Z_{t_1}, Z_{t_2})$, where
	\begin{align*}
		Z_t = \gtruet(\bY) w(\bY; \bm \eta^*) +  \nabla_{\bm \eta^*}^\top \E\{\gtruet(\bY) w(\bY; \bm \eta^*) \}\bm \gamma(\bY, \bX) +  \lambda(\bY, \bX),
	\end{align*}
	and $\lambda(\bY, \bX)$ defined in \Cref{asm_new:s}.
\end{corollary}

\noindent
Asymptotic normality of this estimator for the specific cases of mean and quantile regression was previously established by \citet{Noh2013,Noh2015}.
\citet{Remillard2017} extended the latter to weak convergence uniformly in the quantile level.
Although \citet{Noh2013,Noh2015,Remillard2017} are framed differently, they operate under regularity conditions similar to ours, and lead to the same expression for the asymptotic (co)variances.
Our corollary extend them by allowing for almost arbitrary regression problems with potentially multivariate responses.
A further generalization that allows for discrete variables can be obtained similarly from our main theorems.

Lastly, we can verify the conditions of the theorems for the nonparametric \cref{wex:n} and its bootstrapped version  $	\wboot(\by) = \wt f_{\bY, \bX}(\by, \bx) / \int \wt f_{\bY, \bm X}(\by, \bx) d\bx$, where
\begin{align*}
	\wt f_{\bY, \bm X}(\by, \bx) = \frac{1}{nb_n^d \sigma_n^p}\sum_{i=1}^n \xi_i \prod_{j = 1}^d K\biggr(\frac{Y_{i, j} - y_j}{b_n}\biggr)\prod_{k = 1}^p K\biggl(\frac{X_{i, k} - x_k}{\sigma_n}\biggr).
\end{align*}

\begin{corollary}[Simple kernel estimator] \label{cor:n}
	Under \Cref{asm_new:identifiability,asm_new:euclidean,asm_new:frechet,asm_new:n} with $r_n = \sigma_n^{-p/2}$, the \cref{wex:n} satisfies the conditions of \Cref{thm:consistency,thm:weak_convergence,thm:bootstrap}.
	The bias satisfies
	\begin{align*}
		\bias =  -\frac{\sigma_n^2 \int s^2 K(s) ds}{2} \Vtrue^{-1} \sum_{k = 1}^p \bigl[ \partial_{x_k}^2 \psi(\bx) \times  f_\bX(\bx) +  2\partial_{x_k} \psi(\bx) \times \partial_{x_k} f_\bX(\bx) \bigr]
		+ o(s_n),
	\end{align*}
	where $s_n = n^{-1/2}\sigma_n^{-p/2} + \sigma_n^2$ and $\psi(\bx) = \E\{\gtrue(\bY) \mid \bX = \bx\}$.
	If, additionally, $\gtruet$ is continuous almost everywhere for all $t \in T$, we have
	\begin{align*}
		\cov(\G_{t_1}, \G_{t_1}) = \biggl\{\int K(s)^2 ds\biggr\}^p \E\{ g_{\thtrue, t_1}(\bY)   g_{\thtrue, t_2}(\bY) \mid \bX = \bx\} f_\bX(\bx).
	\end{align*}
\end{corollary}

This estimator is biased, which is expected from kernel based methods.
In the case of mean-regression (i.e., $\gth(y) = y - \theta$), one can verify that the bias and variances are asymptotically equivalent to those of the Nadaraya-Watson estimator \citep[e.g.,][Theorem 3.1]{fan1996local}.
The simple kernel estimator therefore has no advantages over traditional kernel methods and should only be seen as an illustrative example.

Benefits of the copula-based approach can be expected when imposing more structure on the copula model.
For example, \citet{Nagler2016} showed that simplified vine copulas evade the curse of dimensionality prevalent in nonparametric estimation.
Similar effects can be expected with other hierarchical models, like nested Archimedean copulas \citep[see e.g.,][]{okhrin2013structure} or the aggregation copula model of \citet{cote2015}.
This is confirmed empirically in the following section.

%!TEX root = ../copula_ee.tex

\section{Numerical validation}
\label{sec:simulations}
The proposed methodology subsumes a range of problems and estimation methods too wide to be exhaustively covered here.
Instead, we validate our theoretical results in simple settings, using default implementations for all estimators.
We also provide a brief comparison to some benchmark methods.

We simulate from a linear Gaussian model $Y = \bm \beta \bX + Z$, with $\bX \in \R^p$ a vector of \emph{iid} $X \sim N(0, \bm \Omega)$ where $\bm \Omega_{i,j} =  \Ind(i = j) + 0.3 \Ind(i \neq j)$, and $Z \sim N(0,1)$ independent of $X$.
We set $\bm \beta = (1,\dots, 1)^\top / \sqrt p$ to ensure that the signal-to-noise ratio is unaffected by the number of covariates $p$.
The main advantage of this example is that the copula of $(Y,\bX)$ is known to be Gaussian.
As such, we can use the maximum likelihood estimation as baseline, and compare it to our method when the densities are estimated (semi/non)parametrically.

We use one parametric, one semiparametric, and two nonparametric estimators of $\what$.
The first, \emph{Gaussian + Gaussian}, is fully parametric and constructed by fitting Gaussian marginal distributions and a Gaussian copula using  maximum-likelihood at each step.
The second, \emph{KDE + Gaussian}, is semiparametric and uses kernel estimators for the marginal distributions along with the Gaussian copula.
The last two are fully nonparametric, but differ in the copula estimator.
The method \emph{KDE + KDE} uses \cref{wex:n} and \emph{KDE + Kernel Vine} uses a nonparametric vine estimator \citep[\texttt{tll2} in][]{Nagler2017}.

These estimators are provided as the \texttt{R} package \texttt{eecop} \citep{nagler2020a},
providing routines to fit and predict conditional expectiles and quantiles using copulas.
Vine-related functionality is powered by \texttt{rvinecopulib} \citep{nagler2020b}, the Gaussian copula by \texttt{copula} \citep{copula}, and kernel estimators for the margins by \texttt{kde1d} \citep{kde1d}.
The scripts to reproduce the results are in the online supplement.

In \Cref{sec:simul_param_nonparam}, we look at the accuracy and convergence rate of estimators, in \Cref{sec:simul_bootstrap_es} we study the coverage of bootstrapped confidence intervals.
In both cases, we consider both expectile and quantile regression at levels
$t = 0.5$, that is mean and median regression respectively, and $t = 0.95$.
Because the results are qualitatively similar across estimation targets, we only report numbers averaged across targets.
Full results and additional simulations with discrete covariates can be found in the supplementary materials.

\subsection{Estimation accuracy}\label{sec:simul_param_nonparam}

\begin{figure}
  \centering
  \includegraphics[width = 0.9\linewidth]{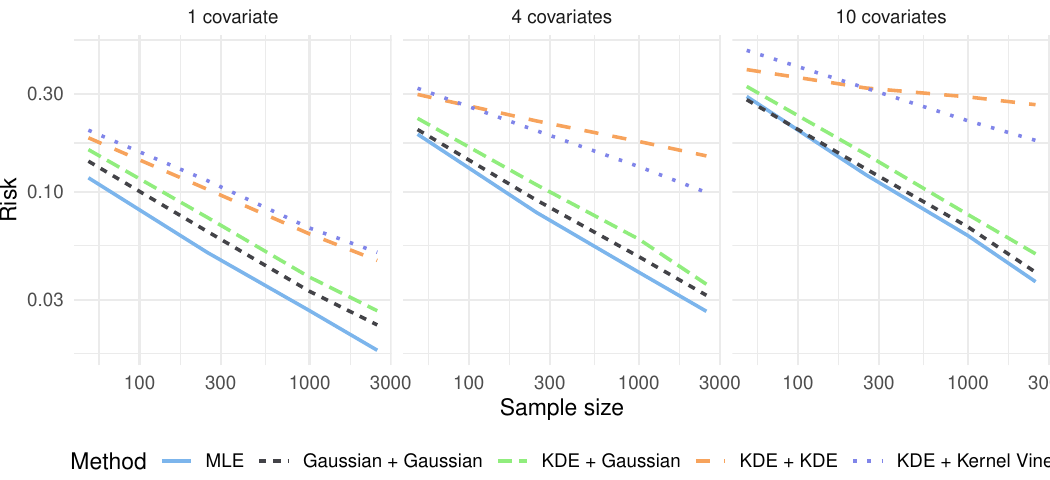}
  \caption{Average risk for expectile and quantile regression in a linear Gaussian model. Both axes have logarithmic scale.}
  \label{fig:param_vs_nonparam}
\end{figure}

We evaluate each estimator's accuracy using its empirical risk.
In other words, we calculate $\left\{50^{-1}\sum_{i=1}^{50} \epsilon_i^2  \left|t - \Ind(\epsilon_i \leq 0)\right|\right\}^{1/2}$ for expectiles and $50^{-1}\sum_{i=1}^{50} \epsilon_i  \{t - \Ind(\epsilon_i \leq 0)\}$ for quantiles, where $\epsilon_i = Y_i - \thhat_{\bX_i, t}$, on an independent test sample $(\bX_1, \bY_1), \dots, (\bX_{50}, \bY_{50})$.
All results are based on 100 replications.
The x-axis of \Cref{fig:param_vs_nonparam} contains the training sample size $n$, the y-axis the risk; both  have logarithmic scale.

There are two main observations.
First, all parametric and semiparametric methods appear to converge at $\sqrt{n}$ rate, as seen from the $-1/2$ slopes.
As expected from \Cref{cor:p,cor:s}, the MLE is slightly more efficient than the parametric copula-based method, which itself is slighlty more efficient than the semiparametric copula-based method.
Second, the convergence rate of the \emph{KDE + KDE} estimator deteriorates with the number of covariates, in line with \Cref{cor:n}.
The \emph{KDE + Kernel Vine} estimator achieves a better rate of convergence, largely unaffected by the dimension, as suggested by our discussion following \Cref{cor:n}.
This illustrates the potential advantage of structural copula models for nonparametric regression.

\subsection{Bootstrap confidence intervals}\label{sec:simul_bootstrap_es}

\begin{figure}
  \centering
  \includegraphics[width = \textwidth]{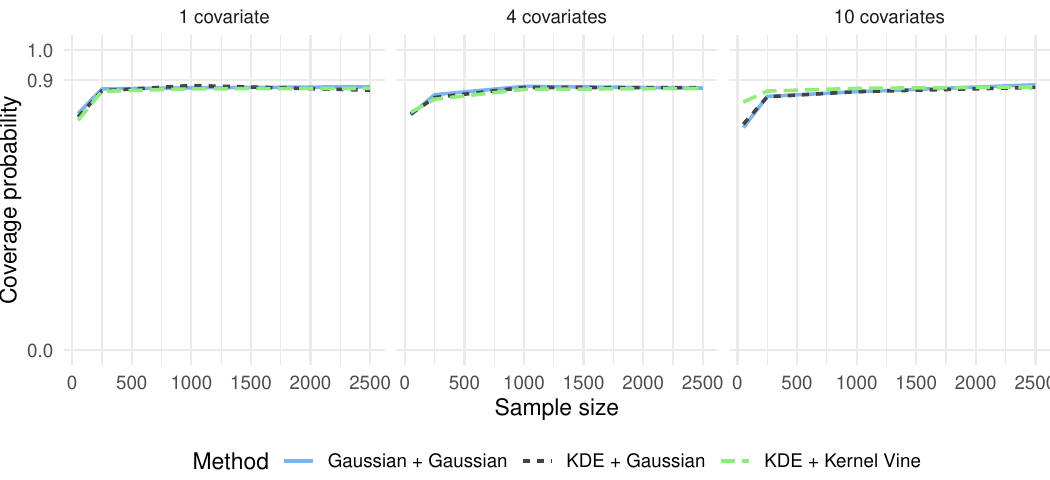}
  \caption{Coverage frequencies of bootstrapped 90\%-confidence intervals.}
  \label{fig:coverage}
\end{figure}

We now consider the same setup, but compute 90\% confidence intervals using 500
bootstrapped replicates.
Because nonparametric estimators are generally biased, we have to undersmooth them to make the bias asymptotically negligible.
In preliminary experiments, we discovered a small bias between the estimated regression target and its bootstrap replicates nevertheless.
More details and a justification are given in the supplementary material.

The method \emph{KDE + KDE} is omitted in these experiments because it
is both heavily biased and computationally too demanding.
In \cref{fig:coverage}, we observe that the coverage probabilities are all close to the target level of 90\%.
Coverage generally improved for larger sample sizes, but is insufficient for very small samples with $n = 50$.
Additional results for all sample sizes and the uncorrected bootstrap are in the supplementary materials.

\subsection{Comparison to other methods} \label{sec:simul_comparison}

We conclude with a comparison of the \emph{KDE + Kernel Vine} estimator from the previous section, \emph{Copula} in the following, to three quantile regression benchmarks:
linear quantile  regression \citep[\emph{LQR},][]{quantreg}, support vector machines \citep[\emph{SVM},][]{kernlab}, and gradient boosted trees \citep[\emph{GBM},][]{gbm}.
All implementations are used with default parameters.

We simulate from two models with $d = 10$ covariates.
The first is the Gaussian from the previous section with marginal distributions of the response and covariates transformed respectively to lognormal and exponential.
The second is a cubic additive model with three active covariates, namely
$Y = (X_1 + X_2 + X_3)^3 + \epsilon$, where $X_1, \dots, X_{10}$ and $\epsilon$ as before.

Using a test sample of size 250, we compare the estimators in two ways.
First, we estimate conditional quantiles at levels 0.5 and 0.95, and report the average absolute distance from the true conditional quantiles.
Second, we estimate conditional quantiles at all levels in $
\{0.05, 0.1, \cdots, 0.9, 0.95\}$, and 
report the fraction of observations where a $\tau_1$-quantile and a $\tau_2$-quantile with a certain gap $\tau_1 - \tau_2$ cross.
All results are based on 100 replications.

\begin{figure}
  \begin{minipage}[t]{0.59\textwidth}
    \begin{figure}[H]
      \includegraphics[width = \textwidth]{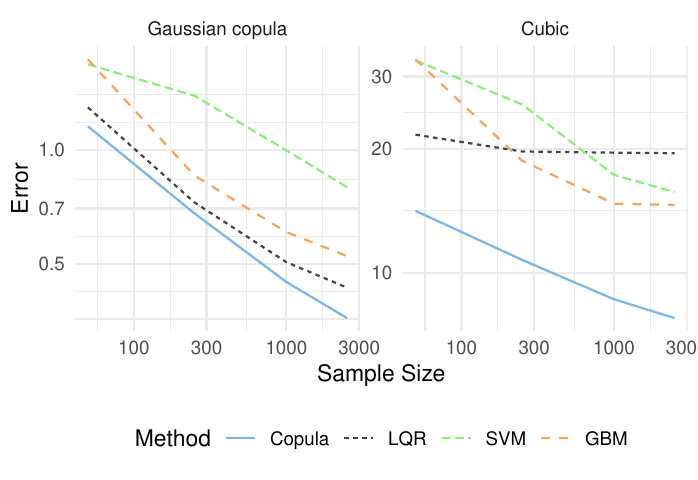}
      \caption{Average absolute test error of the conditional quantiles.}
      \label{fig:benchmark}
    \end{figure}
  \end{minipage} \hfill
  \begin{minipage}[t]{0.37\textwidth}
    \begin{table}[H]
      \caption{\label{tab:crossing_tab}Fraction of test points with quantile crossing.}
      \centering
      \begin{tabular}[t]{>{}r|llll}
      \toprule
      gap & Copula & LQR & SVM & GBM\\
      \midrule
      0.05 & 0\% & 52\% & 42\% & 99\%\\
      0.10 & 0\% & 37\% & 16\% & 75\%\\
      0.30 & 0\% & 27\% & 1\% & 22\%\\
      \bottomrule
      \end{tabular}
      \end{table}
  \end{minipage}
\end{figure}

\Cref{fig:benchmark} indicates that our method outperforms the benchmarks for both models.
This was expected from the Gaussian, which amounts to a copula regression, but not necessarily from the cubic.
\Cref{tab:crossing_tab} further shows that our method satisfy the monotonicity constraints of conditional quantiles, which is not the case for the benchmarks.
% (To account for round-off errors, we only count crossings with a relative difference of more than 0.1\%.)
\emph{GBM} quantiles almost always cross when the gap is 0.05, on 22\% of the samples even at a gap as large as 0.3.
\emph{SVM} and \emph{LQR} do only slightly better. 
Depending on context, such behavior may be prohibitive. 
Quantiles predicted by our method never cross.

%!TEX root = ../copula_ee.tex

\section{Application: quantitative asset allocation} \label{sec:applications}

We revisit the application mentioned in \Cref{sec:intro} and
illustrate how our method can help blend (subjective) economic views in an otherwise quantitative asset allocation.
Suppose that an investor faces the problem of shifting the fraction of her total wealth across various categories to take advantage of evolving market conditions.
Such categories might be asset classes, such as bonds, stocks, and commodities, or industry sectors in a stock portfolio.
She might also need to evaluate a range of performance and risk measures for different portfolios under economic scenarios like ``the GDP will grow by 0.5\%'' or ``the unemployment rate will decrease by 1\%'', over a given horizon.

% Following \cite{markovitz1952}, we use the variance as a measure of portfolio risk t

% Let the mean and covariance matrix of the returns conditionally on $\bX = \bx$ for some $\bx$ be $\bmu = E(\bY \mid \bX = \bx)$ and $\bSigma = \cov(\bY \mid \bX = \bx)$.

Denote by $\bY\in \R^d$, $\bgam \in \R^d$, and $\bX \in \R^p$ the returns on the different categories, the fractions of total wealth invested in each of them, and the covariates.
As measures of portfolio performance and risk, we consider the conditional mean,
standard deviation or \emph{volatility},
and 95\% \emph{Value-at-Risk} \citep[VaR, see e.g., ][]{mcneil2015quantitative},
defined respectively as $\mu_{\bgam}^{*} = \E(\bgam^{\top}\bY \mid \bX = \bx)$,  $\sigma_{\bgam}^{*} = \var(\bgam^{\top}\bY \mid \bX = \bx)^{1/2}$,
and $q_{\bgam}^{*} = F_{-\bgam^{\top}\bY \mid \bX}^{-1}(0.95\mid \bx)$.

Let the indexing set be $T = \{1, 2, 3\} \times \Gamma$, where $\Gamma = \{\bgam \colon \sum_j \gamma_j = 1, \max_{j} |\gamma_j| \le M \}$ for some $M$ representing a constraint on how large a position in a single category can get.
% $\theta_t = \theta_{(j, \bgam)} \in \{\mu_{\bgam},\sigma_{\bgam},q_{\bgam} \}$
% can be identified by the function
% \begin{align*}
%   \gtht = g_{\theta, (j, \bgam)} \in \{\bgam^{\top} \by - \mu_{\bgam},(\bgam^{\top}\by - \mu_{\bgam}) (\bgam^{\top}\by - \mu_{\bgam})^{\top} - \sigma_{\bgam}^2,0.05 - \Ind(\bgam^{\top}\by < q_{\bgam}) \}.
% \end{align*}
The parameter space and identifying functions are
$\Theta = \{\theta_t \colon t \in T\} = \{\theta_{(j, \bgam)} \colon j \in \{1, 2, 3\}, \bgam \in \Gamma \}$ and
$\Gcal = \{\gtht \colon t \in T\} = \{g_{\theta, (j, \bgam)} \colon j \in \{1, 2, 3\}, \bgam \in \Gamma \}$, where
\begin{align*}
	\theta_t = \theta_{(j, \bgam)} =
	\begin{cases}
		\mu_{\bgam}, \quad    & j = 1 \\
		\sigma_{\bgam}, \quad & j = 2 \\
		q_{\bgam}, \quad      & j = 3
	\end{cases}
	\qquad
	\gtht = g_{\theta, (j, \bgam)}  =
	\begin{cases}
		\bgam^{\top} \by - \mu_{\bgam}, \quad                                                            & j = 1 \\
		(\bgam^{\top}\by - \mu_{\bgam}) (\bgam^{\top}\by - \mu_{\bgam})^{\top} - \sigma_{\bgam}^2, \quad & j = 2 \\
		0.95 - \Ind(-\bgam^{\top}\by < q_{\bgam}) , \quad                                                & j = 3
	\end{cases}.
\end{align*}
With $\{\bm e_i\}_{i=1}^d$ the standard basis of $\R^d$, $t = (1, \bm e_i)$ and $ t=(2, \bm e_i)$ correspond respectively to the conditional expected return and variance of asset $i$,
and $t = (2, \bm e_i + \bm e_j)$ to the conditional covariance between assets $i$ and $j$.
This formulation thus also identifies the vector of expected returns $\bmu^* = \E(\bY \mid \bX = \bx)$ and the covariance matrix $\bSigma^* = \cov(\bY \mid \bX = \bx)$.

Given an estimated weight $\what$, $\wh q_{\bgam}$ has to be solved for numerically, but there are closed form solutions for the conditional mean vector and covariance matrix estimators, that is
\begin{align*}
	\wh \bmu =  \frac{\sum_{i = 1}^n\bm Y_i \what(\bY_i)}{\sum_{i = 1}^n \what(\bY_i)}, \qquad \wh \bSigma = \frac{\sum_{i = 1}^n(\bm Y_i - \wh \bmu)(\bm Y_i - \wh \bmu)^\top \what(\bY_i)}{\sum_{i = 1}^n \what(\bY_i)},
\end{align*}
leading to $\wh \mu_{\bgam}= \bgam^{\top}\wh{\bm \mu}$ and $\wh \sigma_{\bgam} = \sqrt{\bgam^{\top}\wh \bSigma \bgam}$.

If for all $\bgam \in \Gamma$, $\btheta_{\bgam} = (\mu_{\bgam}, \sigma_{\bgam}, q_{\bgam})$ lies in a compact subset of $\R \times \R_{>0} \times \R_{>0}$,
the collection of identifying functions $\Gcal$ satisfies the conditions of our theorems.
Thus, our results in \Cref{sec:asymptotics} imply that the estimators $\wh \btheta_{\bgam} = (\wh \mu_{\bgam}, \wh \sigma_{\bgam}, \wh q_{\bgam})$ are consistent, converge weakly, and that their bootstrap versions yield valid inferences, uniformly in $\bgam \in \Gamma$.
To obtain uniform confidence bands over portfolios for the volatility and VaR, we compute the statistics $\wh{\sigma}_{b} = \sup_{\bgam \in \Gamma}\bigl\vert \wh{\sigma}_{b,\bgam} - \wh{\sigma}_{\bgam}\bigr\vert$
and $\wh{q}_{b} = \sup_{\bgam \in \Gamma}\bigl\vert \wh{q}_{b,\bgam} - \wh{q}_{\bgam}\bigr\vert$
for bootstrap samples $b = 1, \dots, B$.
Uniform $\alpha$-confidence bands are then given by $\bgam \mapsto \wh{\sigma}_{\bgam} \pm \wh s_{\sigma,\alpha}$ and $\bgam \mapsto \wh{q}_{\bgam} \pm \wh s_{q,\alpha}$, with $\wh s_{\sigma, \alpha}$ and $\wh s_{q, \alpha}$ respectively the $\alpha$-quantiles of $\wh{\sigma}_{b}$ and $\wh{q}_{b}$.
Scripts to reproduce the following analysis with the \texttt{eecop} \citep{nagler2020a} package are in the supplementary material.

\subsection{The data}

For $\bY$, we use value-weighted returns on 5 industry portfolios.
For $\bX$, we use the real gross domestic product and the seasonally adjusted unemployment rate.
With yearly data covering 1947-2019, we have a total of 72 observations.
The average yearly returns on stocks are in the 13-17\% range, but with large variations over time.
While the GDP has been growing steadily at around 3\% per year, the evolution of the unemployment has varied widely.
Further, the returns on all industry sectors are positively correlated among each other and the growth in GDP, and negatively correlated with the growth in unemployment.
Because the auto-correlations of all variables and their squares are statistically indistinguishable from zero, we treat the data as \emph{iid}.
Plots of time-series, auto-correlation functions, cross-correlations and summary statistics, along with additional details on the data sources, are in the supplementary material.

To study the impact of the predictors in the quantitative allocation scheme, we create three scenarios for the change in GDP and unemployment: good economy (+4.35/-11.52), median economy (+3.22/-2.48), poor economy (+1.59/+5.44).
The good and bad scenarios are obtained using the 75th and 25th percentile for the growth in GDP unemployment, and conversely for the unemployment.

\subsection{Results}

To derive predictions for each scenario, we estimate all margins by kernel estimators with plug-in bandwidths and fit a parametric vine copula model for the copula density.
Predicted conditional means, standard deviations, and 95\% VaRs are given in \Cref{tab:appli_tab2b}.
We see that, the better the economic outlook, the higher the expected returns, and the lower the risk.

\begin{table}

	\caption{\label{tab:appli_tab2b}Predicted means ($\mu$), standard deviations ($\sigma$), and VaRs ($q$) for each scenario.}
	\centering
	\resizebox{\linewidth}{!}{
		\begin{tabular}[t]{lrrrrrrrrrrrrrrr}
			\toprule
			\multicolumn{1}{c}{ } & \multicolumn{3}{c}{Cnsmr} & \multicolumn{3}{c}{Manuf} & \multicolumn{3}{c}{HiTec} & \multicolumn{3}{c}{Hlth} & \multicolumn{3}{c}{Other}                                                                                      \\
			\cmidrule(l{3pt}r{3pt}){2-4} \cmidrule(l{3pt}r{3pt}){5-7} \cmidrule(l{3pt}r{3pt}){8-10} \cmidrule(l{3pt}r{3pt}){11-13} \cmidrule(l{3pt}r{3pt}){14-16}
			Scenario              & $\mu$                     & $\sigma$                  & $q$                       & $\mu$                    & $\sigma$                  & $q$  & $\mu$ & $\sigma$ & $q$  & $\mu$ & $\sigma$ & $q$  & $\mu$ & $\sigma$ & $q$  \\
			\midrule
			Good economy          & 17.7                      & 15.0                      & 3.7                       & 17.5                     & 13.1                      & 7.2  & 17.5  & 18.2     & 11.8 & 17.9  & 18.6     & 12.1 & 18.4  & 16.5     & 9.2  \\
			Median economy        & 16.1                      & 15.9                      & 5.4                       & 14.7                     & 13.4                      & 11.1 & 15.6  & 18.9     & 16.6 & 16.2  & 18.3     & 12.2 & 15.2  & 16.7     & 9.8  \\
			Poor economy          & 9.1                       & 16.6                      & 13.2                      & 8.0                      & 14.2                      & 12.8 & 9.5   & 20.1     & 25.0 & 11.1  & 17.6     & 17.3 & 7.1   & 17.0     & 18.8 \\
			\bottomrule
		\end{tabular}}
\end{table}

\begin{figure}[h!]
	\centering
	\includegraphics[width = 0.9\linewidth]{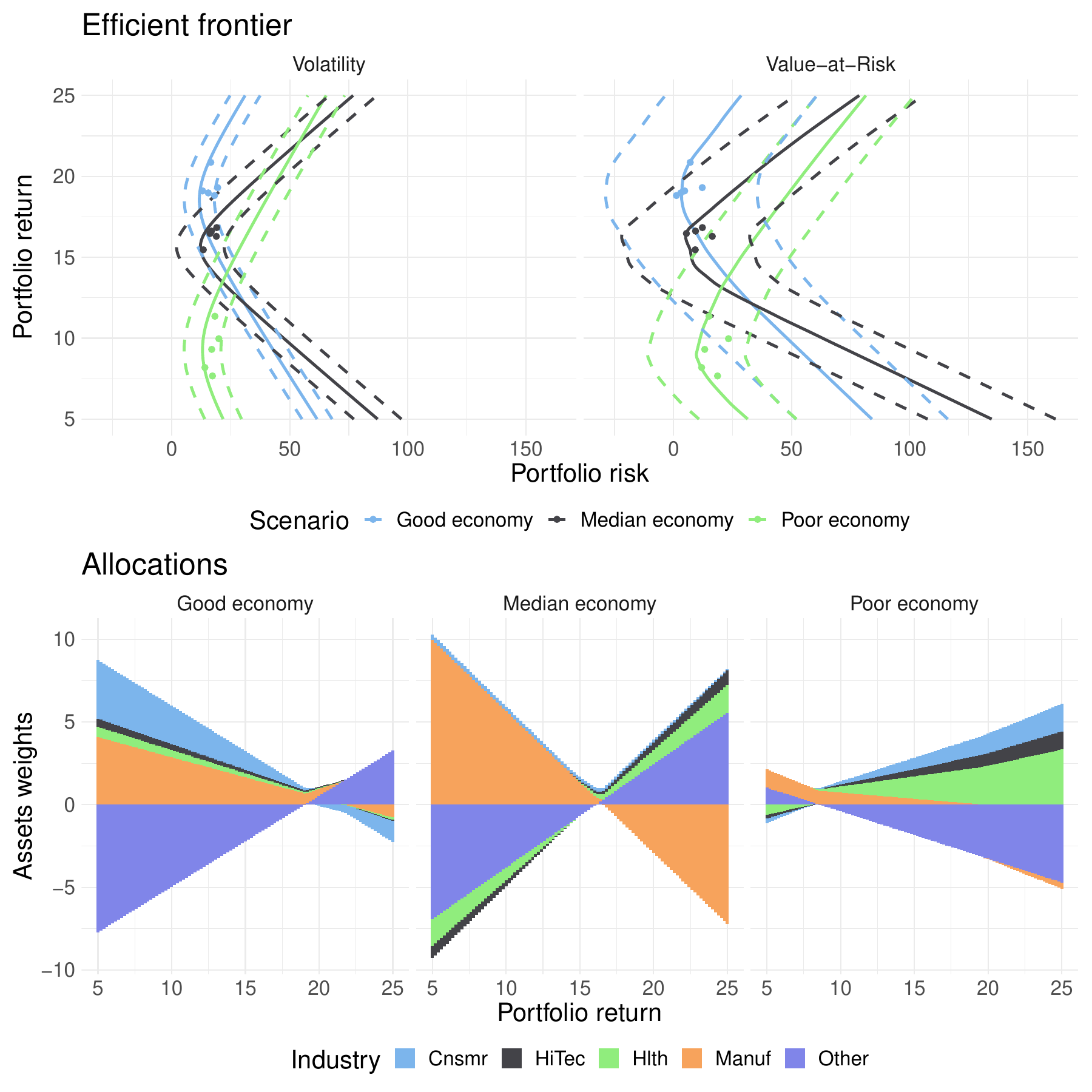}
	\caption{Efficient frontiers with uniform 75\%-confidence bands and scatterplots for the individual assets (top panel), and corresponding portfolio weights (bottom panel).}
	\label{fig:appli_fig3}
	\vspace{1cm}
\end{figure}

For each scenario, we compute the \emph{efficient frontier}, that is a set of portfolios where $\bgam$ can be expressed as a linear combination of the minimum variance and market portfolios.
Background details on asset allocation can be found in the supplementary material.
In the top panel of \Cref{fig:appli_fig3}, we show the
frontiers as a lineplot with uniform 75\% bands as the dashed lines, and the corresponding quantities for the individual assets as a scatterplot.
In the bottom panel, the portfolio weights are displayed as a function of the expected
return.
We can make the following observations:
\begin{itemize}
	%   \item For each scenario, the MV and MKT portfolios can be identified in the top-left pannel
	%   as respectively the left-most point and tangency point of the lines passing through the origin to the corresponding parabola.
	\item The upper half of all the parabolas yields higher returns for the same risk compared with the lower half,
	      rational investors would only choose such portfolios.
	\item Similarly, since the scatterpoints corresponding to the individual assets are ``inside'' the parabolas,
	      rational investors would prefer diversified portfolios from the upper half.
	\item Better economic outlooks generally imply higher expected returns at a given risk level.
	\item  The weights of efficient portfolios with returns between that of minimum variance and market portfolios are generally reasonable, that is mostly included in $[-1, 1]$.
	      And targeting returns outside this range generally requires large long and short positions, accompanied by quick an increase in portfolio risk.
\end{itemize}
Nonetheless, the uncertainties are fairly large,
due to the small sample size.
In higher frequency data  (e.g., monthly) however, serial dependence can no longer be ignored.
An extension of our results to this setting is discussed briefly in the following section.

\section{Discussion}
\label{sec:conclusion}

\Cref{sec:asymptotics} provides an umbrella theory for solutions to copula-based estimating equations,
and has close connections to several recent results.
\citet{Noh2013} shows consistency and asymptotic normality for mean regression in a semiparametric model.
\citet{Noh2015} and \citet{Debacker2017} derive similar results for quantile regression, the latter for semiparametric copula densities.
\citet{Remillard2017} establish weak convergence of parametetric and semiparametric estimators and validity of a parametric bootstrap procedure for the conditional quantile function indexed by the level.
Note that \citet{Noh2015} replace the \emph{iid}-assumption by a mixing condition, which we do not cover.
But our results can likely be extended to stationary sequences under more stringent conditions using the techniques in \citet{Dehling2002}.
Nonetheless, we generalize and extend the above in several ways:
\begin{itemize}
	\item The focus of previous research is on mean and quantile regression with univariate response.
	      We allow for large classes of potentially multivariate identifying functions.
	      This opens new possibilities for copula-based solutions to other regression problems.
	\item Previous results cover only parametric or semiparametric copula estimators.
	      Our theory allows for parametric, semiparametric, and nonparametric methods.
	\item Weak convergence of $\thhat$ as a process indexed by some potentially dense set $T$ is established.
	      It can be used to derive simultaneous and coherent inferences across arbitrary combinations of finite/infinite-dimensional features of the conditional distribution.
	\item We propose and validate a bootstrapping scheme.
	      It is applicable whenever obtaining the asymptotic distribution in closed form is inconvenient or infeasible.
	\item The theory also applies to M-estimators, defined as maximizers of a criterion function, when the criterion is differentiable \citep[see, e.g.,][Section 2.2.6]{Kosorok2007}.
	\item Our results apply to both continuous and discrete data.
	      Admittedly, the practical applicability for discrete data is limited by a lack of available software.
	      While experimental features for discrete variables exist in the \texttt{rvinecopulib} package \citep{nagler2020b}, they still need to mature.
	      But the theoretical foundation is set and we hope that the computational limitations are overcome in the near future.
\end{itemize}
Being broadly applicable, our main results are somewhat abstract.
But it is often straightforward to specialize them given a specific identifying function and weight estimator, as we do in \Crefrange{cor:p}{cor:n}.
We close our discussion by outlining applications to additional regression problems covered by our theory in \Cref{sec:asymptotics}.

First, suppose the goal is to characterize the relationship between a response $Y_1$ and a treatment $Y_2$ using an instrument $Y_3$, conditionally on a set of exogenous covariates $\bX$.
Specifically, assume as in \citet{Newey2003} that
$Y_1 = \bm b(Y_2)^\top \bm \btheta(\bX) + Z$,
where $\bm b$ is known vector of basis functions, and $Z$ is a zero-mean error term.
When the treatment is endogenous, that is $Y_2 \not \independent Z$, identifying $\btheta$ requires an instrument $Y_3$ satisfying
$E(Z\mid \bX = \bx, Y_3 = y_3) = 0$
for all $\bx$ and $y_3$.
In other words, one has to solve
$\E \bigl( Y_1 \vert \bX = \bx, Y_3 = y_3 \bigr) = E\bigl\{ \bm b(Y_2)^\top \bm \btheta(\bX) \vert \bX = \bx, Y_3 = y_3 \bigr\}$ for all $\bx$ and $y_3$.
Then the collection
$\Gcal = \{  g_{\btheta, t} \colon \btheta \in \Theta, t \in T\}$ with $g_{\btheta, t}(\by)= b_t(y_3) \bigl\{ y_1 - \bm b(y_2)^\top \btheta  \bigr\}$ identifies the parameter $\bm \theta(\bm x)$ for given $\bm x$.
In the supplement, we show that \Cref{asm_new:euclidean,asm_new:frechet} are satisfied, and it is then left to check \Crefrange{asm_new:wn}{asm_new:h_functions_cont} and \Cref{asm_new:boot}.

Further, our results apply to problems with censored or missing responses by tweaking the weight function.
For example, suppose we only observe the right-censored version $\bar S = \min(S, Z) \in \R$ of a survival time $S$, along with a censoring indicator $\Delta = \Ind(S \le Z)$.
For any identifying function $\phi_\theta$,
$\E\bigl\{ \phi_{\theta} (S)   \mid \bX = \bx \bigr\} = 0$ if and only if
$E\bigl\{ \phi_{\theta} (\bar S)  \Delta \bar w^*(\bar S) \zeta(\bar S) \bigr\} = 0$,
where $\zeta(t) = 1 /\{1 - F_{Z\mid \bX}(t \mid  \bx)\}$ and $\bar w^*$ is defined as in~\eqref{eq:w_x}, but replacing  $F_{Y, \bX}$ and $F_{Y}$ by $F_{\bar S, \bX}$ and $F_{\bar S}$.
This technique was used in \citet{Debacker2017} to allow for right censoring in copula-based quantile regression,
and it fits the setup of \Cref{sec:asymptotics}.
With $\bY = (\bar S, \Delta)$, $g_\theta(\by) = \phi_\theta(y_1) y_2$, and $\wtrue(\by) = \bar w^*(y_1) \zeta(y_1)$, then
$
	E\bigl\{ \phi_{\theta} (\bar S)  \Delta \bar w^*(\bar S) \zeta(\bar S) \bigr\} = E\bigl\{ g_{\theta} (\bY)  \wtrue(\bY) \bigr\}
$
and it remains to check the assumptions of \Crefrange{thm:consistency}{thm:bootstrap}.
The above is only one instance of a larger class of methods based on \emph{inverse probability weighting}.
Other weight functionals $\zeta$ can be used similarly to account for other forms of censoring and missingness \citep[see, e.g.][]{Robins1994, Woolridge2007, Han2016}.
Following an earlier version of this paper that appeared online, this idea was already picked by \cite{hamori2020copula}, albeit in a restricted semiparametric context.

\if1\blind{
\subsubsection*{Acknowledgements}

The authors thank Johannes Wiesel for the proof of Lemma 11 in the supplementary material.  We are grateful to the Associate Editor and three referees for helpful comments. Part of the research was conducted while Thomas Nagler was at Delft University of Technology and Thibault Vatter was at Columbia University.

\subsubsection*{Funding}

Thomas Nagler was partially supported by a Spinoza grant awarded by the Netherlands Organisation of Scientific Research (NWO). Thibault Vatter was partially supported by the Swiss National Science Foundation  (Grant 174709).
} \fi

\appendix
\section{Assumptions}\label{sec:asm}%

In general, we assume that the parameter space satisfies $\Theta \subseteq \ell^\infty(T)$ for some indexing set $T$.
For simplicity, we consider $T$ to be a compact subset of a Euclidean space.
We also use $(\bY', \bX')$ to denote an independent copy of $(\bY, \bX)$.

\subsection{Assumptions for the main results}\label{sec:asm:main}

\begin{assumption}\label{asm_new:identifiability}
	The parameter space $\Theta$ is compact and contains an interior point $\thtrue \in \Theta$ such that $E\{\gtrue(\bY)\wtrue(\bY)\} = 0$ and, for any $\epsilon > 0$, $\inf_{|\theta_t - \thtruet| > \epsilon, t\in T} \vert E\{\gtht(\bY)\wtrue(\bY)\} \vert > 0$.
\end{assumption}

\begin{assumption}\label{asm_new:euclidean}
	The class $\Gcal = \{\gth\colon \theta \in \Theta\}$ is Euclidean\footnote{A formal definition is given in the supplementary material.} for some envelope function $G$.
	% That is, there exists finite positive constants $A$ and $V$ such that, for every probability measure $Q$ with $0 < \|G\|_{L_1(Q)} < \infty$, 
	%     \begin{align*}
	%         N(\epsilon \|G\|_{L_1(Q)}, \Gcal, L_1(Q)) \leq A \epsilon^{-V},\qquad 0 < \epsilon \leq 1,
	%     \end{align*}
	%     where $A$ and $V$ do not depend on $Q$.

\end{assumption}

\begin{assumption}\label{asm_new:wn}
	There exists a sequence of functions $w_{n}\colon \R^{2d + p} \to \R$ and a sequence $r_n = o(n^{1/2})$ with $r_n \ge 1$, such that
	$\max_{1 \le i \le n} \biggl\vert \what(\bY_i) - \frac{1}{n} \sum_{j=1}^n w_{n}(\bY_i,\bY_j,\bX_j)  \biggr\vert = o_P\bigl(n^{-1/2}r_n\bigr)$.
\end{assumption}
\begin{assumption} \label{asm_new:consistency}
	$\sup_{\theta \in  \Theta} \| \E\{\gth(\bY) w_n(\bY, \bY', \bX')\} -  \E\{\gth(\bY) \wtrue(\bY)\} \|_T  = o(1)$.
\end{assumption}

\begin{assumption}\label{asm_new:g_wn}
	$G$, $w_{n}$, and $r_n$ are such that
	\begin{subassumption}
		\item\label[assumption]{asm_new:g_wn:alpha_moment}
		$\E\{ \vert G(\bY) \vert \} < \infty$,
		\item\label[assumption]{asm_new:g_wn:gwn1}
		$\E\{G(\bY) |w_n(\bY, \bY, \bX)| \} = o(n^{1/2}r_n)$,
		\item\label[assumption]{asm_new:g_wn:gwn2}
		$\E\{G(\bY)^2 w_n(\bY, \bY', \bX')^2 \} = o(nr_n^2)$.
	\end{subassumption}
\end{assumption}

\begin{assumption}\label{asm_new:h_functions}

	\setlength{\belowdisplayskip}{0pt} \setlength{\belowdisplayshortskip}{0pt}
	\setlength{\abovedisplayskip}{0pt} \setlength{\abovedisplayshortskip}{0pt}

	With $(\by', \bx')$ an arbitrary point in $\Ycal \times \Xcal$,
	the functions
	\begin{align*}
		\htht(\by', \bx') & = \E\{\gtht(\bY)w_n(\bY, \by', \bx') + \gtht(\by')w_n(\by', \bY, \bX)\}, \\
		H_{n}(\by', \bx') & = \E\{G(\bY)|w_n(\bY, \by', \bx')| + G(\by')|w_n(\by', \bY, \bX)|\},
	\end{align*}
	satisfy
	\begin{subassumption}
		\item\label[assumption]{asm_new:h_functions:equicont}
		$\sup_{\|\theta_1 - \theta_2 \|_{T} + \|t_1 - t_2\| < \delta_n}\E\{|h_{n, \theta_1,t_1}(\bY', \bX') - h_{n,\theta_2,t_2}(\bY', \bX')|^2\} = o(r_n^2)$ for every $\delta_n \to 0$,
		\item\label[assumption]{asm_new:h_functions:mom}
		$\E\{r_n^{-2}H_n(\bY', \bX')^{2}\} = O(1)$,
		\item\label[assumption]{asm_new:h_functions:lindeberg}
		$\E\{r_n^{-2}H_n(\bY', \bX')^{2}\Ind_{r_n^{-1} H_n(\bY', \bX') \geq \eta \sqrt{n}}   \} = o(1)$ for every $\eta > 0$.
	\end{subassumption}
\end{assumption}

\begin{assumption}\label{asm_new:h_functions_cont}
	$\| \E[\{g_{\theta_n}(\bY) - \gtrue(\bY)\}\{w_n(\bY, \bY', \bX') - \wtrue(\bY)\}] \|_T = o(\|\theta_n - \thtrue\|_T)$ for every $\theta_n \to \thtrue$.
\end{assumption}

\begin{assumption}\label{asm_new:frechet}
	The map $\theta \mapsto \E\{\gth(\bY)\wtrue(\bY)\}$ from $\ell^\infty(T)$ to $  \ell^\infty(T)$ is Fr\'echet differentiable in a neighborhood of $\thtrue$ and the derivative $V_{\theta}\colon \ell^{\infty}(T) \mapsto \ell^{\infty}(T)$ is invertible at $\thtrue$.
	That is, for $\theta$ in a neighborhood of $\thtrue$,  $V_{\theta}$ is a bounded linear operator such that
	\begin{align*}
		\| \E\{g_{\theta_n}(\bY)\wtrue(\bY)\} - \E\{\gth(\bY)\wtrue(\bY)\} - V_{\theta}(\theta_n - \theta) \|_T = o(\| \theta_n - \theta \|_{T}), \quad \forall \theta_n \to \theta.
	\end{align*}
	And the inverse is the map $\Vtrue^{-1}$ such that $\Vtrue^{-1}\Vtrue$ is the identity.
\end{assumption}

\begin{assumption}\label{asm_new:boot}
	There are \emph{iid} random variables $\xi_{i}, i = 1, \dots, n$ independent of $(\bY_i, \bX_i)_{i = 1}^n$ with $\E(\xi_{1}) = \var(\xi_{1}) = 1$, and  $\E\{|\xi_{1}|^{2 + \epsilon}\} < \infty$ for some $\epsilon < 0$, such that for the same $w_n,  r_n$ as in \Cref{asm_new:wn},
	$\max_{1 \le i \le n} \biggl\vert \wboot(\bY_i) - \frac{1}{n }  \sum_{j=1}^n \xi_i  w_{n}(\bY_i,\bY_j,\bX_j)  \biggr\vert = o_P\bigl(n^{-1/2}r_n\bigr)$.
\end{assumption}

A brief discussion of the assumptions is in order.
\Cref{asm_new:identifiability} ensures identifiability of the parameter of interest $\thtrue$.
\Cref{asm_new:euclidean} limits the complexity of the class of identifying functions $\Gcal = \{g_\theta\colon \theta \in \Theta\}$.
Importantly, this complexity is disentangled from the weight estimator.
Euclidean classes \citep{nolan1987} generalize Vapnik-Cervonenkis classes of real-valued functions and are also called VC-type classes by some authors \citep[e.g.,][]{gine2006}.
A formal definition and several convenient properties of these classes are in the supplementary material, where we also verify \Cref{asm_new:euclidean} for all examples from \Cref{sec:examples_identifying}.
\Cref{asm_new:wn} allows us to expand $\what$ as a sample average and a uniformly negligible remainder.
Since we only evaluate $\what$ on $\bY_1, \dots, \bY_n$ in the empirical estimating equation \eqref{eq:ee_estimator}, the expansion only needs to be valid at these points.
\Cref{asm_new:boot} is an analogous condition for $\wboot$,  the bootstrap version of $\what$.
\Cref{asm_new:consistency} ensures consistency of $\what$ to $\wtrue$ in a weak sense.
In particular, we require neither uniform nor pointwise consistency, although,  with \Cref{asm_new:g_wn}, either can be shown to be sufficient.

The remaining assumptions deal with joint regularity of the class of identifying functions $\Gcal$ and the weight estimator $\what$ through the sequence $w_n$.
\Cref{asm_new:g_wn} and \Crefrange{asm_new:h_functions:mom}{asm_new:h_functions:lindeberg} are moment conditions.
In the latter two, the randomness from either estimating $\wtrue$ or replacing the population mean by its empirical counterpart has been averaged out; see also the discussion following \Cref{thm:weak_convergence}.
\Cref{asm_new:h_functions:equicont} is a form of stochastic equicontinuity in both $\theta$ and $t$.
Note that \Crefrange{asm_new:h_functions:equicont}{asm_new:h_functions:lindeberg} are common in the context of function classes changing with $n$ \cite[see e.g.,][Section 2.11.3]{vanderVaart1996}.
\Cref{asm_new:frechet} ensures sufficient smoothness of the map  $\theta \mapsto \E\{g_\theta(\bY) \wtrue(\bY)\}$ and is verified in the supplementary material for the examples from \Cref{sec:examples_identifying}.
It could be weakened to Hadamard differentiability at the cost of slightly more tedious proofs.
\Cref{asm_new:h_functions_cont} ensures that replacing $\wtrue$ by $\what$ has a negligible effect on the smoothness in \Cref{asm_new:frechet}.

When specifying $\what$ further, the assumptions can often be disentangled into separate conditions on the identifying functions and weight estimator, see \Crefrange{asm_new:p}{asm_new:n} below.

\subsection{Assumptions for the corollaries}\label{sec:asm:cor}

\begin{assumption}\label{asm_new:p}
	Denote respectively by $\wtrue (\by) = w(\by; \bm \eta^*)$ and $\what (\by) = w(\by; \bm \ethat)$ the true and estimated weight functions.
	\begin{subassumption}
		\item \label[assumption]{asm_new:p:mle} One has
		$\bm \ethat- \bm \eta^*
			=    \frac 1 n \sum_{i=1}^n \bm \gamma(\bY_i, \bX_i)  + o_P(n^{-1/2})$
		and $\bm \etboot- \bm \eta^*
			=     \frac 1 n \sum_{i=1}^n \xi_i \bm \gamma(\bY_i, \bX_i)  + o_P(n^{-1/2})$,
		for a function $\bm \gamma$
		with $\E\{\bm \gamma(\bY, \bX)\} = 0$ and  $\E\{\| \bm \gamma(\bY, \bX)\|_2^{2 + \epsilon} \} < \infty$ and \emph{iid} positive random variables $\xi_{1}, \dots, \xi_{n}$ independent of $(\bY_i, \bX_i)_{i = 1}^n$ with $\E(\xi_{1}) = \var(\xi_{1}) = 1$ and  $\E\{|\xi_{1}|^{2 + \epsilon}\} < \infty$ for some $\epsilon < 0$.
		\item \label[assumption]{asm_new:p:derivs} The function $w(\by; \bm \eta^*)$ is twice continuously differentiable in $\bm \eta^*$ with derivatives uniformly bounded in $\by \in \Ycal$.
		\item \label[assumption]{asm_new:p:moms} One has $\E\{w(\bY; \bm \eta^*)^2 + \| \nabla_{\bm \eta} w(\bY; \bm \eta^*) \|_2^2 \} < \infty$.
		\item \label[assumption]{asm_new:p:G} For some $\epsilon > 0$, one has $\E\{G(\bY)^{2 + \epsilon} \mid \bX = \bx\} < \infty$ and $\E\{G(\bY)^{2 + \epsilon}\}$.
		\item \label[assumption]{asm_new:p:g_cont} For every $\delta_n \to 0$ and with $\| \cdot \|$ the Euclidean norm, one has
		\begin{align*}
			 & \sup_{\|\theta_1 - \theta_2 \|_{T} + \|t_1 - t_2\| < \delta_n}\E\bigl\{|g_{\theta_1, t_1}(\bY) - g_{\theta_2, t_2}(\bY)|^2 \mid \bX = \bx \bigr\} = o(1),    \\
			 & \sup_{\|\theta_1 - \theta_2 \|_{T} + \|t_1 - t_2\| < \delta_n}\bigl\|\E\bigl\{|g_{\theta_1, t_1}(\bY) - g_{\theta_2, t_2}(\bY)|^2  \bigr\} \bigr\|_2 = o(1).
		\end{align*}
	\end{subassumption}
\end{assumption}

\begin{assumption}\label{asm_new:s}
	Write $\bF_{\bY}(\by) = (F_{Y_1}(y_1), \dots, F_{Y_d}(y_d))$ and similarly for $\bF_{\bX}$, $\wb F_{\bY}$, and $\wb F_{\bX}$.
	Define $\omega(\bu, \bv; \bm \eta) =  c_{\bY, \bX}(\bu, \bv; \bm \eta) / c_{\bY}(\bu; \bm \eta)$, $w(\by; \bm \eta) = \omega\{\bF_\bY(\by), \bF_\bX(\bx); \bm \eta\}$, and
	\begin{align*}
		\lambda(\bY, \bX) & =  \sum_{j = 1}^d \int \gtrue(\by)  \frac{\partial \omega\{\bm F_\bY(\by), \bm F_\bX(\bx); \bm \eta^*\}}{\partial F_{Y_j}(y_j)} \times  \{\Ind(Y_{j} \le y_{j}) - F_{Y_j}(y_j)\} f_\bY(\by) d\by   \\
		& \quad +    \sum_{k = 1}^p \frac{\partial  \E\bigl\{ \gtrue(\bY) \mid \bX = \bx \bigr\} }{\partial x_k} \times \frac {\{\Ind(X_k \le x_{k}) - F_{X_k}(x_k)\}} {f_{X_k}(x_k)}  .
	\end{align*}
	Further, for any $Z \in \{Y_1, \dots, Y_d, X_1, \dots, X_p\}$, write
	$\wt F_{Z}(z) = n^{-1} \sum_{i = 1}^n \xi_i\Ind(Z_{i} \le z)$ for the bootstrapped empirical distribution.

	\begin{subassumption}
		\item \label[assumption]{asm_new:s:common} \Cref{asm_new:p:mle,asm_new:p:G,asm_new:p:g_cont} hold.
		\item \label[assumption]{asm_new:s:derivs} The function $(\bm u, \bm v, \bm \eta) \mapsto \omega(\bm u, \bm v; \bm \eta)$ has two continuous derivatives on the domain $[0, 1]^d \times  \{\bm v \colon \| \bm v - \bF_\bX(\bx)\| < \epsilon\} \times \{\bm \eta \colon \| \bm \eta - \bm \eta^*\| < \epsilon\}$ for some $\epsilon > 0$.
		\item \label[assumption]{asm_new:s:moms} One has $\E\{\omega(\bY, \bX; \bm \eta^*)^2 + \| \nabla_{(\bm \eta, \bY, \bX)} \omega(\bY, \bX; \bm \eta^*) \|_2^2  \} < \infty$.
	\end{subassumption}
\end{assumption}

\begin{assumption}\label{asm_new:n}
	{\color{white}l}
	\begin{subassumption}
		\item \label[assumption]{asm_new:n:kernel} $K$ is a symmetric, bounded probability density function on $[- 1, 1]$.
		\item \label[assumption]{asm_new:n:bw} One has $b_n, \sigma_n \to 0$, $b_n^2 = o(n^{-1/2} \sigma_n^{-p /2})$, $\sigma_n^{-p/2}b_n^{d} / \ln n \to \infty$, and $n b_n^d \sigma_n^p / \ln n \to \infty$.
		\item \label[assumption]{asm_new:n:fyx} The densities $f_{\bm X, \bY}$ and $f_\bY$ have uniformly bounded and continuous derivatives up to the third order and $\sup_{\by \in \Ycal} f_{\bm X \mid \bY}(\bx \mid \by) < \infty $.
		\item \label[assumption]{asm_new:n:fy} One has
		$\lim_{\delta \to 0}\sup_{\by \in \Ycal} \sup_{\|\by' - \by\| \le \delta} \sup_{\|\bx' - \bx\| \le \delta}   \biggl\vert \frac{f_{\bY, \bm X}(\by', \bx')}{f_{\bY, \bm X}(\by, \bx) } - 1\biggr\vert = 0$.
		\item \label[assumption]{asm_new:n:G} For some $\epsilon, \delta > 0$, one has $\E\{\sup_{|\bm a| < \delta} G(\bY + \bm a)^{2 + \epsilon} \mid \bX = \bx\} < \infty$ and $\E\{G(\bY)^{2 + \epsilon}\}$.
		\item \label[assumption]{asm_new:n:g_cont} For every $\theta_n \to \thtrue$ and $\delta_n \to 0$ and with $\| \cdot \|$ the Euclidean norm, one has
		\begin{align*}
			\bigl\| \E\bigl\{ |g_{\theta_n}(\bY ) - \gtrue(\bY)| \mid \bX = \bx \bigr\} \bigr\|_T                                                             & = O(\|\theta_n - \thtrue \|_T), \\
			\sup_{\|\theta_1 - \theta_2 \|_{T} + \|t_1 - t_2\| < \delta_n}\E\bigl\{|g_{\theta_1, t_1}(\bY) - g_{\theta_2, t_2}(\bY)|^2 \mid \bX = \bx \bigr\} & = o(1).
		\end{align*}
	\end{subassumption}
\end{assumption}

In all three cases, the specialized assumptions \Crefrange{asm_new:p}{asm_new:n} are standard regularity conditions on the estimation method, as well as smoothness and moment conditions for the identifying function.
This is in contrast to the general conditions \Cref{asm_new:wn,asm_new:g_wn,asm_new:consistency,asm_new:h_functions,asm_new:h_functions_cont,asm_new:boot}, where the estimation method and identifying function are often intertwined.
Unfortunately, to the best of our knowledge, this disentanglement is not feasible in the general setting of \Crefrange{thm:consistency}{thm:bootstrap} without strengthening the conditions.

\bibliography{copula_ee} 

\includepdf[pages=-]{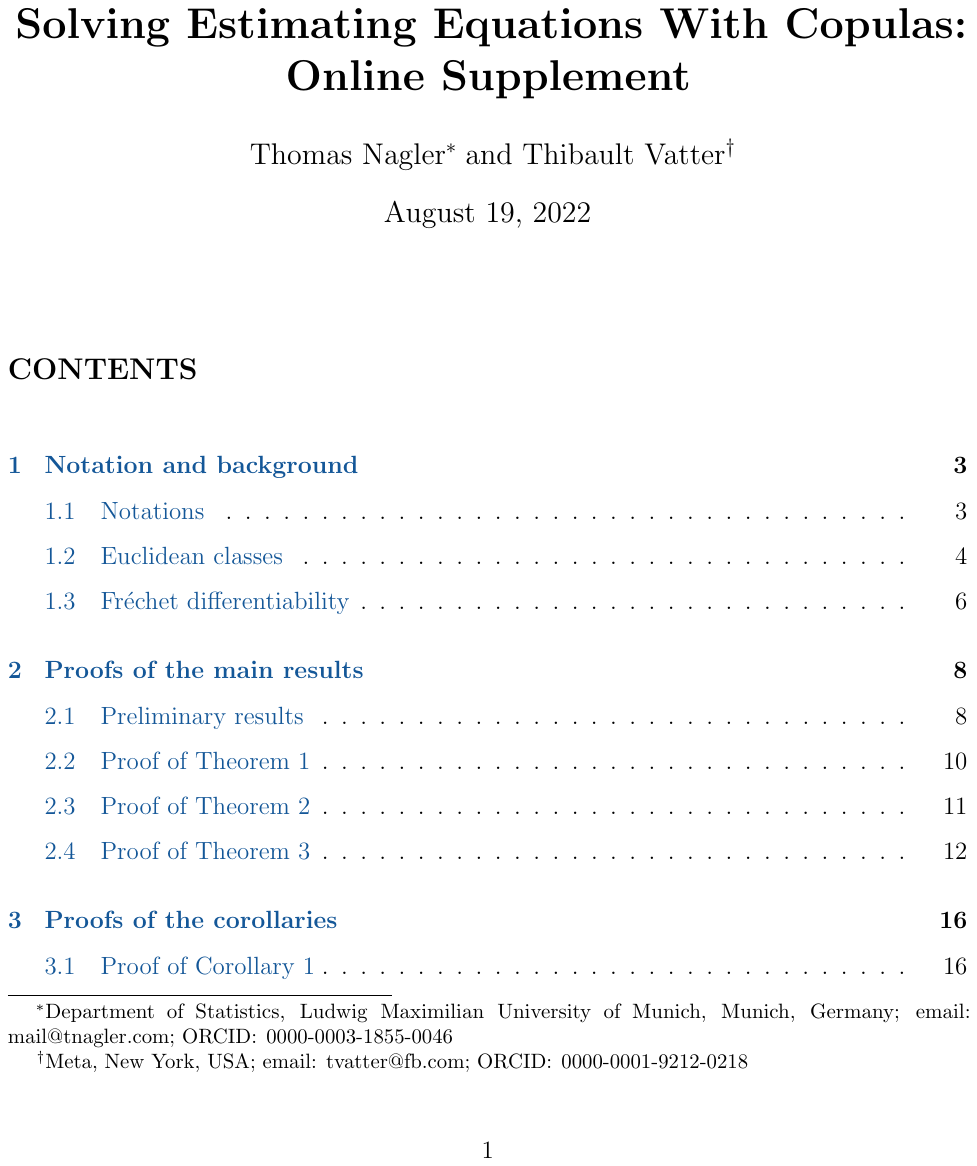}

\end{document}